\documentclass{aa}

\usepackage{graphicx}
\usepackage{CJK}
\usepackage{enumitem}
\usepackage[varg]{txfonts}
\usepackage{amsmath}
\usepackage{amssymb}
\usepackage{footmisc}
\usepackage{booktabs}
\usepackage[colorlinks = true,
            linkcolor = blue,
            urlcolor  = blue,
            citecolor = blue,
            anchorcolor = blue,
            unicode]{hyperref}

\begin{document}

\begin{CJK*}{UTF8}{gbsn}

\title{Evolved Massive Stars at Low-metallicity\\ \uppercase\expandafter{\romannumeral3}. A Source Catalog for the Large Magellanic Cloud}
\titlerunning{Evolved Massive Stars at low-Z \uppercase\expandafter{\romannumeral3}. A Source Catalog for the LMC}

\author{
Ming Yang (杨明) \inst{1} \and Alceste Z. Bonanos \inst{1} \and Biwei Jiang (姜碧沩) \inst{2} \and Jian Gao (高健) \inst{2} \and Panagiotis Gavras \inst{3} \and Grigoris Maravelias \inst{1,4} \and Shu Wang (王舒) \inst{5} \and Xiao-Dian Chen (陈孝钿) \inst{5} \and Man I Lam (林敏仪) \inst{6} \and Yi Ren (任逸) \inst{2} \and Frank Tramper \inst{1} \and Zoi T. Spetsieri \inst{1}
}
\authorrunning{Yang, Bonanos \& Jiang et al.}

\institute{
IAASARS, National Observatory of Athens, Vas. Pavlou and I. Metaxa, Penteli 15236, Greece\\
                \email{myang@noa.gr} \and
Department of Astronomy, Beijing Normal University, Beijing 100875, People's Republic of China \and
Rhea Group for ESA/ESAC, Camino bajo del Castillo, s/n, Urbanizacion Villafranca del Castillo, Villanueva de la Cañada, 28692 Madrid, Spain \and
Institute of Astrophysics, Foundation for Research and Technology-Hellas, Heraklion 71110, Greece \and
Key Laboratory of Optical Astronomy, National Astronomical Observatories, Chinese Academy of Sciences, Datun Road 20A, Beijing 100101, People's Republic of China \and
Key Laboratory of Space Astronomy and Technology, National Astronomical Observatories, Chinese Academy of Sciences, Beijing 100101, People's Republic of China
}

\abstract{
We present a clean, magnitude-limited (IRAC1 or WISE1$\leq$15.0 mag) multiwavelength source catalog for the Large Magellanic Cloud (LMC). The catalog was built upon crossmatching ($1''$) and deblending ($3''$) between the Spitzer Enhanced Imaging Products (SEIP) source list and Gaia Data Release 2 (DR2), with strict constraints on the Gaia astrometric solution to remove the foreground contamination. It is estimated that about 99.5\% of the targets in our catalog are most likely genuine members of the LMC. The catalog contains 197,004 targets in 52 different bands including 2 ultraviolet, 21 optical, and 29 infrared bands. Additional information about radial velocities and spectral/photometric classifications were collected from the literature. We compare our sample with sample from \citet{Gaia2018b}, indicating that the bright end of our sample is mostly comprised of blue helium-burning stars (BHeBs) and red HeBs with inevitable contamination of main sequence stars at the blue end. After applying modified magnitude and color cuts based on previous studies, we identify and rank 2,974 red supergiant, 508 yellow supergiant, and 4,786 blue supergiant candidates in the LMC in six color-magnitude diagrams (CMDs). The comparison between the CMDs from the two catalogs of the LMC and Small Magellanic Cloud (SMC) indicates that the most distinct difference appears at the bright red end of the optical and near-infrared CMDs, where the cool evolved stars (e.g., RSGs, AGB, and RGs) are located, which is likely due to the effect of metallicity and SFH. Further quantitative comparison of colors of massive star candidates in equal absolute magnitude bins suggests that, there is basically no difference for the BSG candidates, but large discrepancy for the RSG candidates as LMC targets are redder than the SMC ones, which may be due to the combined effect of metallicity on both spectral type and mass-loss rate, and also the age effect. The effective temperatures ($T_{\rm eff}$) of massive star populations are also derived from reddening-free color of $(J-K_{\rm S})_0$. The $T_{\rm eff}$ ranges are $3500<T_{\rm eff}<5000$ K for RSG population, $5000<T_{\rm eff}<8000$ K for YSG population, and $T_{\rm eff}>8000$ K for BSG population, with larger uncertainties towards the hotter stars.
}


\maketitle

\section{Introduction}

The satellite galaxies of Milky Way (MW), the Large and Small Magellanic Clouds (LMC and SMC) are two irregular dwarf galaxies visible from the southern hemisphere. The MCs are a unique laboratory for studying stellar populations, star formation, chemical evolution, and so on, as their close distances allow studies of individual stars within them. One of the hot topics for the MCs is the massive star population (initial masses $\gtrsim8~M_\sun$), since they are related to many extreme events in the Universe, for example, supernovae (SN), gravitational waves, black holes, and long gamma-ray bursts \citep{Woosley2002, Smartt2009, Maeder2012, Massey2013, Smith2014, Adams2017, Crowther2019}. The research on massive stars has achieved large progress for the past decades from both photometry and spectroscopy (e.g., \citealt{Humphreys1984, Massey2003, Levesque2006, Evans2011, Yang2011, Yang2012, Davies2013, Gonzalez2015, Yang2018, Neugent2020}). However, the foreground contamination from the MW makes the study of extragalactic massive stars challenging. For example, Galactic dwarfs or giants may have similar brightnesses and/or colors as the background massive stars in the MCs. This problem has been largely mitigated after the release of Gaia Data Release 2 (DR2) data \citep{Gaia2016a, Gaia2016b, Gaia2018a}, since the majority of the foreground contamination can be robustly removed by using the Gaia astrometric solution.

We have already built a source catalog for the SMC \citep{Yang2019}, by applying strict constraints on the Gaia astrometric solution and collecting a large amount of multiwavelength photometry and time-series data, as well as auxiliary spectroscopic and/or photometric classifications from the literature. The massive star candidates, e.g., blue supergiant stars (BSGs), yellow supergiant stars (YSGs), and red supergiant stars (RSGs), were identified by utilizing evolutionary tracks and synthetic photometry. It also showed that there was a relatively clear separation between RSGs and asymptotic giant branch stars (AGBs) down to the tip of the red giant branch (TRGB) after the astrometric constraints (see Figure 15 of \citealt{Yang2019}), even though the following study of \citet{Yang2020} also indicated that the boundary between RSGs and AGBs was still blurred in all aspects. 

In order to fully understand the effect of metallicity, star formation history, binary fraction, etc., on the evolution and physical properties of massive star populations, it is important to expand the similar work to the LMC and beyond. In this paper, we study the evolved dusty massive star populations in the LMC, by establishing a clean source catalog of massive stars. The paper is structured as follows: the multiwavelength source catalog is presented in \textsection2. The identification of evolved massive star candidates is described in \textsection3. \textsection4 is the comparison of the LMC and SMC. The summary is given in \textsection5.  

\section{Multiwavelength Source Catalog}

\begin{figure}
\center
\includegraphics[bb=125 365 455 690, scale=0.65]{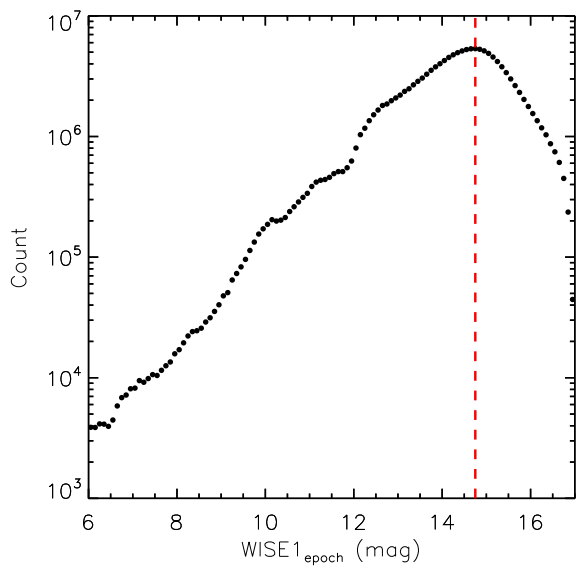}
\caption{Histogram of 144,831,057 ALLWISE WISE1 single-epoch measurements. A drop-off around 14.75 mag is shown by the red dashed line. 
\label{w1_epoch_histo}}
\end{figure}

The procedure of data reduction is the same as \citet{Yang2019}, hence we briefly describe the procedure here. More details can be found in the original paper. The source catalog was built up based on the crossmatching ($1''$) and deblending ($3''$) between Spitzer Enhanced Imaging Products (SEIP) source list and Gaia DR2 \citet{Gaia2016a, Gaia2018a}. SEIP source list includes 12 bands data from near-infrared (NIR) to mid-infrared (MIR) retrieved from Two Micron All Sky Survey (2MASS; \citealt{Skrutskie2006}), Spitzer \citep{Werner2004} and Wide-field Infrared Survey Explorer (WISE; \citealt{Wright2010}). The data cover the main body of the LMC ($64^\circ\leq$ R.A. $\leq 94^\circ$, $-74^\circ\leq$ Decl. $\leq -63^\circ$) with a magnitude cut of IRAC1 or WISE1$\leq$15.0 mag, since there is a drop-off around 14.75 mag in the number counts for 144,831,057 ALLWISE WISE1 single-epoch measurements as shown in Figure~\ref{w1_epoch_histo}. This initial step resulted in 264,292 targets.

\begin{figure*}
\includegraphics[bb=120 375 485 695, scale=0.46]{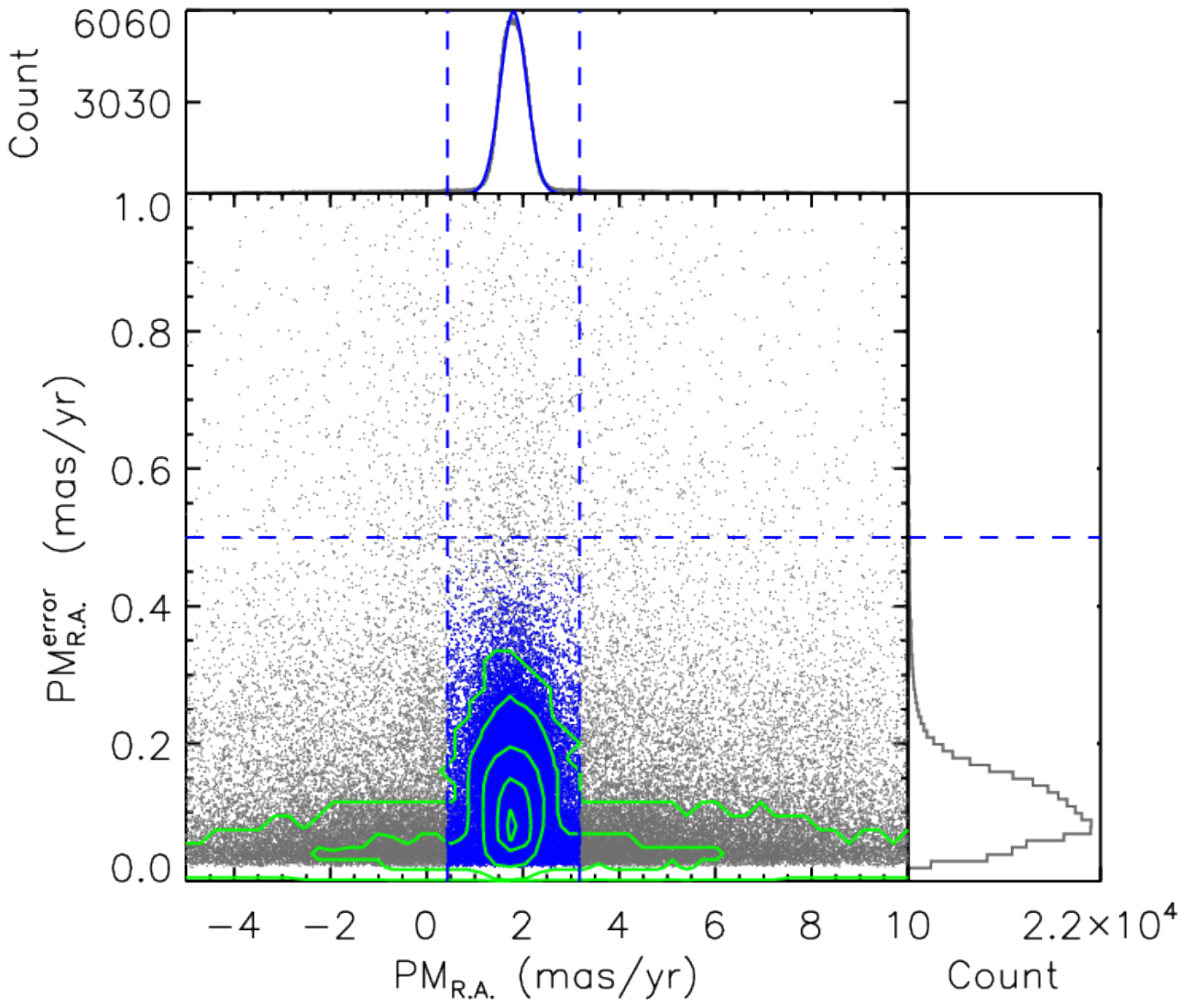}
\includegraphics[bb=120 375 485 695, scale=0.46]{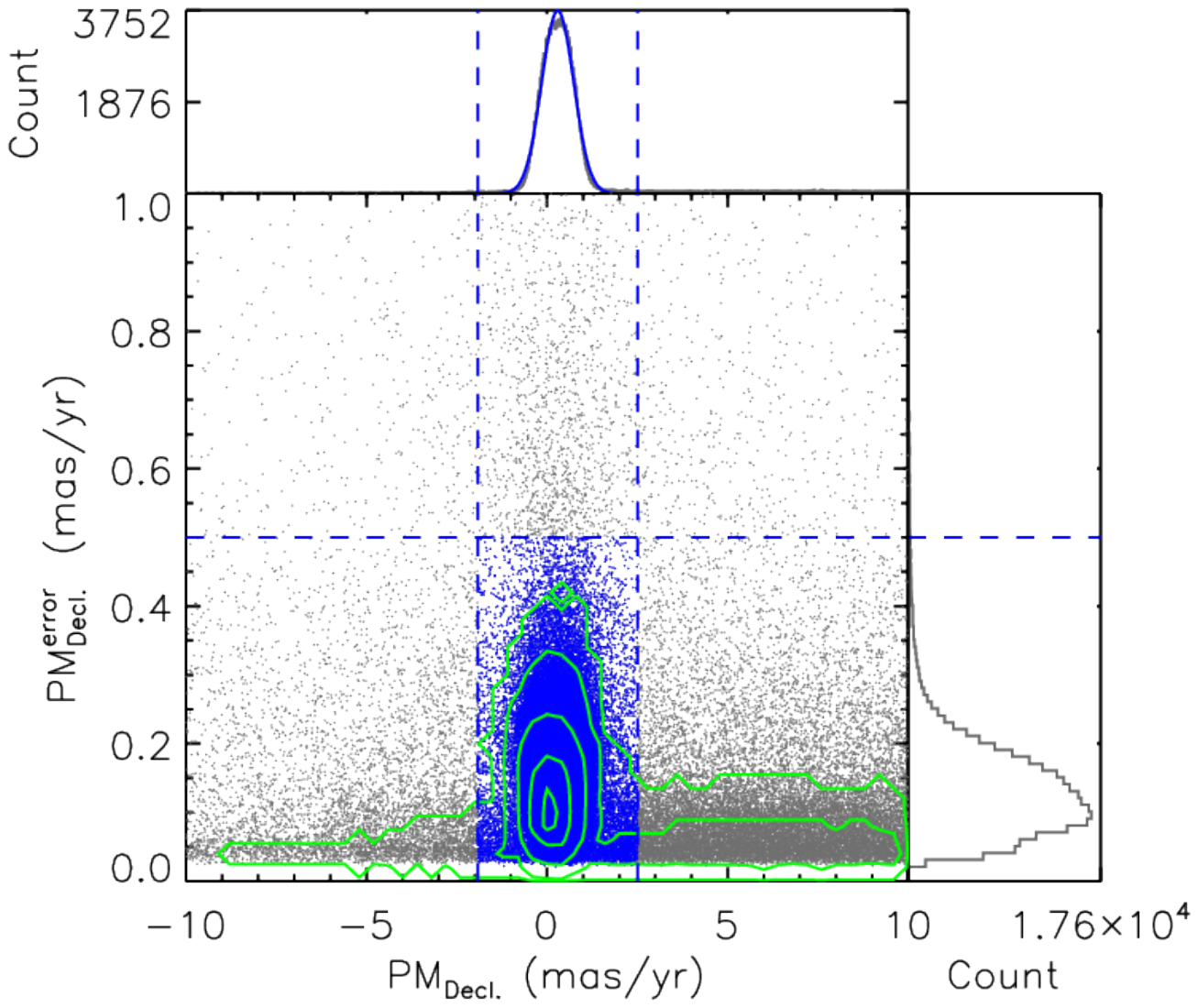}
\includegraphics[bb=100 375 485 695, scale=0.46]{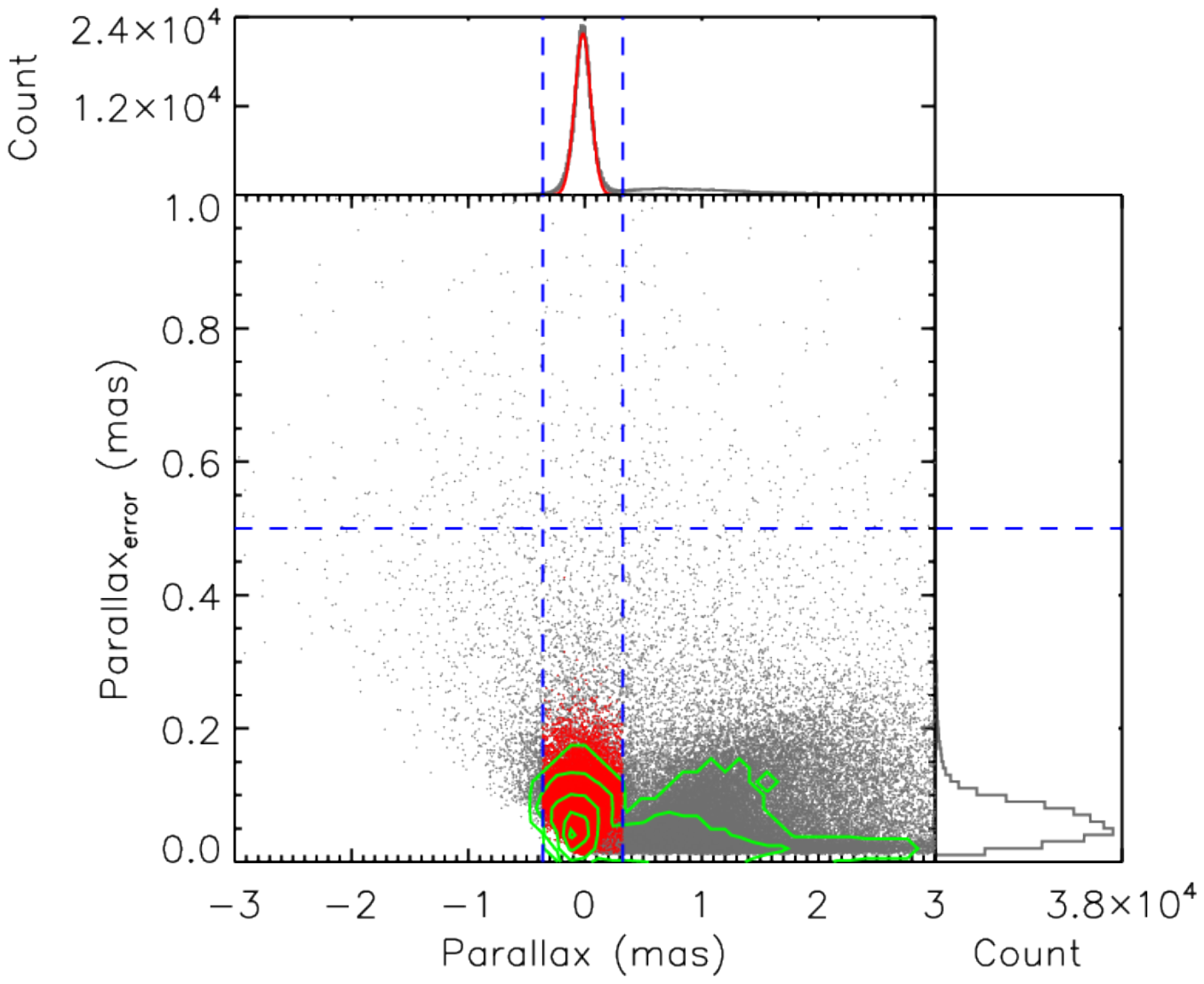}
\caption{Evaluation of the Gaia astrometric solution. Horizontal dashed lines indicate the limits of errors in PMs (0.5 mas/yr) and parallax (0.5 mas). From left to right show Gaia PMs in R.A., Decl, and parallax versus their errors. Gaussian profiles are fitting in each panel and $\pm5\sigma$ are calculated (vertical dashed lines). For final selected targets (red) based on the parallax, except adopted Gaussian fitting, an additional elliptical constraint is also applied with the $5\sigma$ limits of Gaussian profiles in $PM_{\rm R.A.}$ and $PM_{\rm Decl.}$ (based on blue targets in the first two panel) taken as the primary and secondary radii, respectively. Green contours show the number density in each digram. 
\label{pm_constrain}}
\end{figure*}

\begin{figure}
\includegraphics[bb=100 430 475 615, scale=0.67]{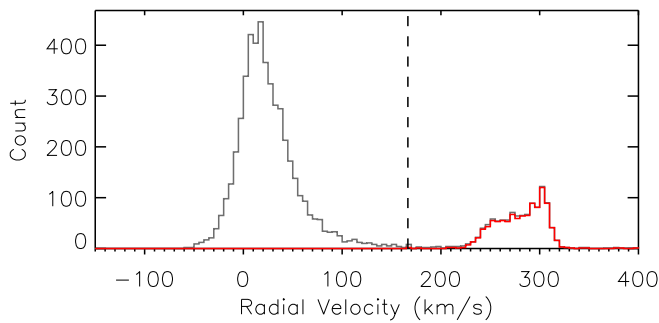}
\caption{Histogram of Gaia RVs. Milky Way and LMC are clearly separated. Selected targets (red) has a minimal value of $\sim$166.5 km/s (dashed line). 
\label{gaia_rv}}
\end{figure}

\begin{figure}
\includegraphics[bb=115 365 465 690, scale=0.67]{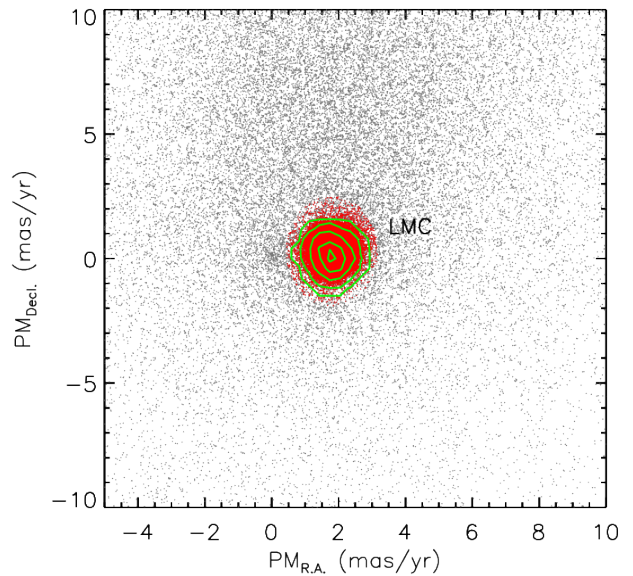}
\caption{$PM_{\rm R.A.}$ versus $PM_{\rm Decl.}$ diagram, for which the contamination of remaining foreground sources for the LMC is estimated around 0.08\% ($\sim$157/197,004) and can be ignored. Red dots indicate the astrometry constrained targets and green contours represent their number density. 
\label{pm_radec}}
\end{figure}

\begin{figure}
\includegraphics[bb=125 365 495 690, scale=0.67]{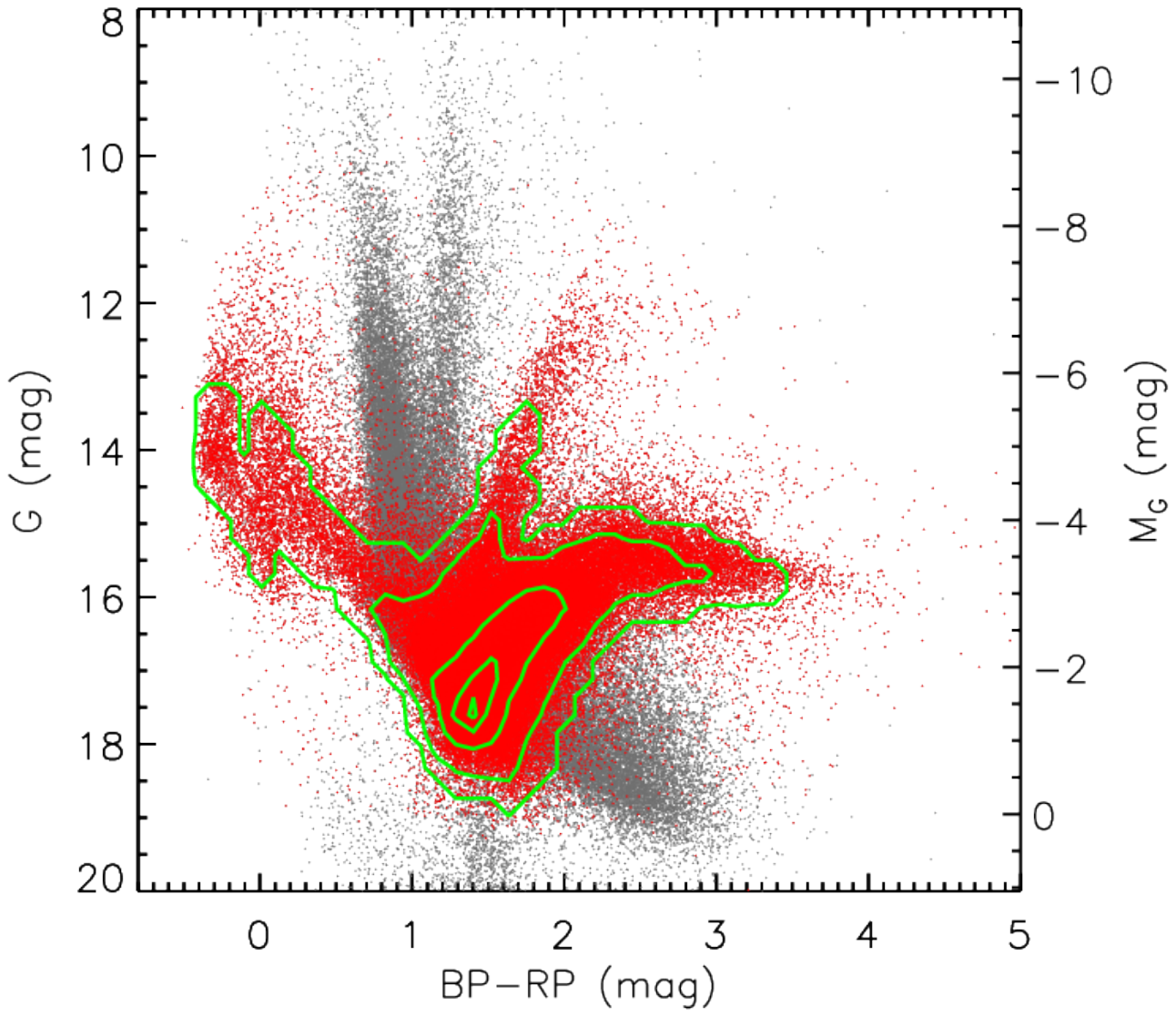}
\caption{Gaia G versus BP-RP diagram before (gray) and after (red) the astrometric constraints. A large number of foreground contamination is cleaned. Green contours show the number density of astrometry constrained targets.
\label{gaia_cmd}}
\end{figure}

The membership of the LMC was then determined by using the Gaia DR2 astrometric solution \citep{Lindegren2018}. Three Gaussian profiles, $1.805\pm0.274$ mas/yr for $PM_{\rm R.A.}$, $0.294\pm0.442$ mas/yr for $PM_{\rm Decl.}$, and $-0.018\pm0.068$ mas for parallax were fitted to targets with errors $<$0.5 mas/yr in proper motions (PMs) and $<$0.5 mas in parallax as shown in Figure~\ref{pm_constrain}. Notice that, for the parallax, except the Gaussian profile, an additional elliptical constraint also applied, where the primary and secondary radii of the ellipse were the $5\sigma$ limits of Gaussian profiles in $PM_{\rm R.A.}$ and $PM_{\rm Decl.}$, respectively (see Figure~\ref{pm_radec}). Moreover, a constraint on the radial velocity (RV) $\geq166.5$ km/s was also applied for all targets having Gaia RV measurements as shown in Figure~\ref{gaia_rv}. In total, this resulted in 197,005 targets. We note that, by visually inspecting the Gaia color-magnitude diagram (CMD; see Figure~\ref{gaia_cmd}), there was an obvious outlier in the upper right region, which turned out to be a foreground star with $RV=39$ km/s. Still, 1 out of 197,005 is an extreme small rate of contamination. After removing this target we had a final sample of 197,004 sources as shown in Figure~\ref{gaia_cmd}, where a true distance modulus of $18.493\pm0.055$ mag for the LMC was adopted (same below; \citealt{Pietrzynski2013}). Notice that, for convenience, the extinction correction was not applied when converting visual magnitudes into absolute magnitudes, since foreground extinction was relatively small at the line of sight of the LMC ($A_{\rm V}\approx0.2$ mag, if $E(B-V)\sim0.06$ mag with the Galactic average value of $R_{\rm V}=3.1$ was adopted; \citealt{Oestreicher1995, Dobashi2008, Gao2013}) and neither the internal extinction structure of the LMC nor the distance of the targets were accurately determined. In that sense, the internal extinction of the LMC might largely vary from star to star, especially for targets close to the star formation region (e.g. massive stars). Hence, the converted absolute magnitude only represents a lower limit for each individual star.

Further evaluation of astrometric excess noise, which measures the disagreement and is expressed as an angle between the observations of a source and the best-fitting standard astrometric model (using five astrometric parameters), indicated that 98.17\% and 99.78\% targets had $astrometric\_excess\_noise\leq0.5$ and $\leq1.0$ mas, respectively. Moreover, \citet{Gaia2018b} provided lists of possible members based on the analysis of Gaia PMs and parallaxes for 75 Galactic globular clusters, nine dwarf spheroidal galaxies, one ultra-faint system, and the MCs. Their basic idea was to first determine the median and robust scatter in PMs and parallaxes by selecting a sample of stars covering a larger field of view, then further eliminate any sources showing larger scatter in the PMs, and finally construct a filter based on a covariance matrix of the cleaned sample, which allows one to properly select out likely members. The comparison between our sample and \citet{Gaia2018b} indicated that 99.98\% of our targets were consistent with their results. This suggests that our results are highly reliable without sophisticated correction (see more details and discussion in Section 2 of \citealt{Yang2019}).

Figure~\ref{pm_radec} shows $PM_{\rm R.A.}$ versus $PM_{\rm Decl.}$. Based on this diagram, we estimated that the contamination of the remaining foreground and the possible non-point/background sources (SEIP sources without valid 2MASS measurements) for the LMC to be around 0.08\% ($\sim$157/197,004) and 0.5\% ($\sim$1,034/197,004), respectively, and could be ignored. In that sense, about 99.5\% of our targets are most likely genuine members of the LMC. 

\begin{figure*}
\center
\includegraphics*[bb=50 395 540 660, scale=0.9]{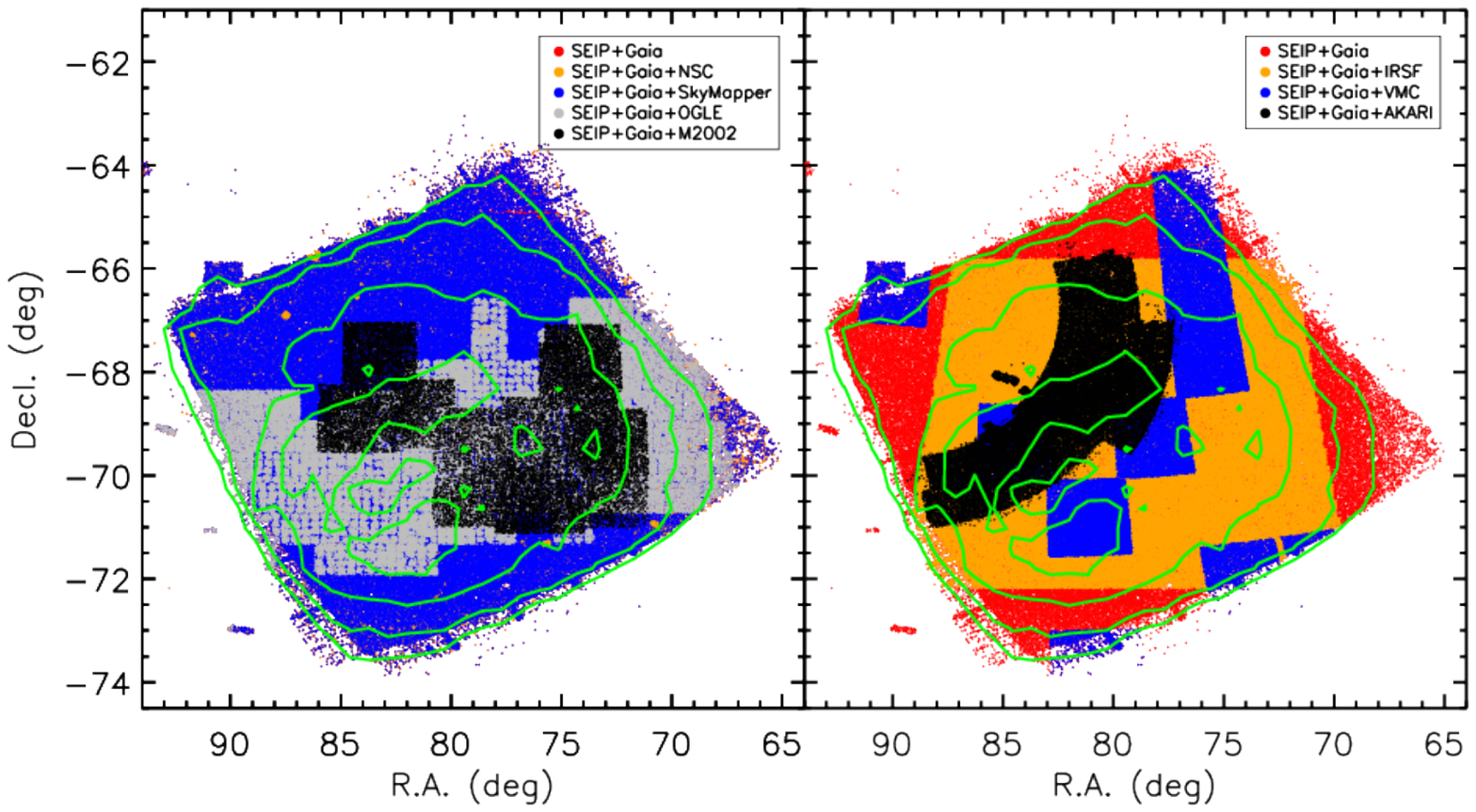}
\caption{Spatial distribution of the optical (left) and IR (right) datasets.  
\label{spatial}}
\end{figure*}

After selecting the fiducial sample of 197,004 targets, we retrieved additional data (deblended with a search radius of $3''$) by using a search radius of $1''$ from different dataset ranging from ultraviolet (UV) to far-infrared (FIR) as shown below (more details about each dataset can be found in \citealt{Yang2019}):
\begin{itemize}[noitemsep,topsep=0pt,parsep=0pt,partopsep=0pt] 
	\item 44,524 matches ($\sim$23\%) from the VISTA survey of the Magellanic Clouds system (VMC) DR4 \citep{Cioni2011}).
    \item 136,935 matches ($\sim$70\%) from the IRSF Magellanic Clouds point source catalog (MCPS) \citep{Kato2007}.
	\item 37,116 matches ($\sim$19\%) from AKARI LMC point source catalog \citep{Onaka2007, Murakami2007, Kato2012}.
    \item 49 matches from HERschel Inventory of the Agents of Galaxy Evolution (HERITAGE) band-merged  source catalog (units are in flux [mJy] instead of magnitude; \citealt{Pilbratt2010, Meixner2013, Seale2014}).
    \item 162,462 matches ($\sim$82\%) from SkyMapper DR1.1 \citep{Keller2007, Bessell2011, Wolf2018}. 
	\item 151,476 matches ($\sim$77\%) from the NOAO source catalog (NSC) DR1 \citep{Nidever2018}). 
	\item 20,379 matches ($\sim$10\%) from a UBVR CCD survey of the Magellanic Clouds by \citet{Massey2002} (M2002).
	\item 74,104 matches ($\sim$38\%) from Optical Gravitational Lensing Experiment (OGLE; \citealt{Udalski1992, Szymanski2005, Udalski2008, Udalski2015}) Shallow Survey in the LMC \citep{Ulaczyk2012}.
	\item 215 matches from revised GALEX source catalog for the All-Sky Imaging Survey (GUVcat\_AIS) \citep{Morrissey2007, Bianchi2017}.
\end{itemize} 
In total, there are 52 filters including 2 UV, 21 optical, and 29 IR filters. The spatial distributions of the additional optical (left) and IR (right) datasets are shown in Figure~\ref{spatial} (GALEX and HERITAGE data are not shown in the diagram since the matches are scarce).

Additional classifications were also retrieved from the literature with a search radius of $1''$, including:
\begin{itemize}[noitemsep,topsep=0pt,parsep=0pt,partopsep=0pt] 
	\item 113 matches from \citet{Whitney2008}, for which $\sim$1000 young stellar objects (YSOs) and also some evolved stars were identified in the LMC based on their IR color and spectral energy distribution (SED). 
	\item 143,447 matches from \citet{Boyer2011}, who investigated the IR properties of cool, evolved stars in the MCs using observations from Spitzer.
	\item 124 matches from \citet{Jones2017}, where nearly 800 point sources observed by Spitzer Infrared Spectrograph (IRS; \citealt{Houck2004}) were classified using a decision tree method based on their infrared spectral features, continuum and SED shape, bolometric luminosity, cluster membership and variability information. 
	\item 98 matches from \citet{Massey2003}, who used optical spectra to identify hundreds of RSGs in the MCs based on photometric data from \citet{Massey2002}.
	\item 620 matches from \citet{Bonanos2009}, which is a catalog for 1,750 massive stars in the LMC from the literature with accurate spectral types, and a multiwavelength photometric catalog for a subset of 1,268 of these stars, with the goal of exploring their infrared properties.
	\item 63 matches from \citet{Neugent2012}, who investigated the evolution of YSGs and RSGs in the LMC by identifying them based on the optical spectroscopy and comparing them with the new Geneva evolutionary models on the CMD.
	\item 108 matches from \citet{Evans2015}, for which 263 massive stars in the north-eastern region of the LMC were spectrally classified.
	\item 156 matches from \citet{Gonzalez2015}, for which physical properties of about 500 RSGs in the LMC and SMC were studied by using NIR/MIR photometry and optical spectroscopy. 
	\item 31,919 matches from Simbad \citep{Wenger2000}. We retrieved RVs, optical spectral classifications, main object types, and auxiliary object types.
\end{itemize}
Unmatched targets are most likely due to larger PMs, blends, or the photometric quality cuts. 

\begin{figure*}
\center
\includegraphics[bb=55 365 560 710]{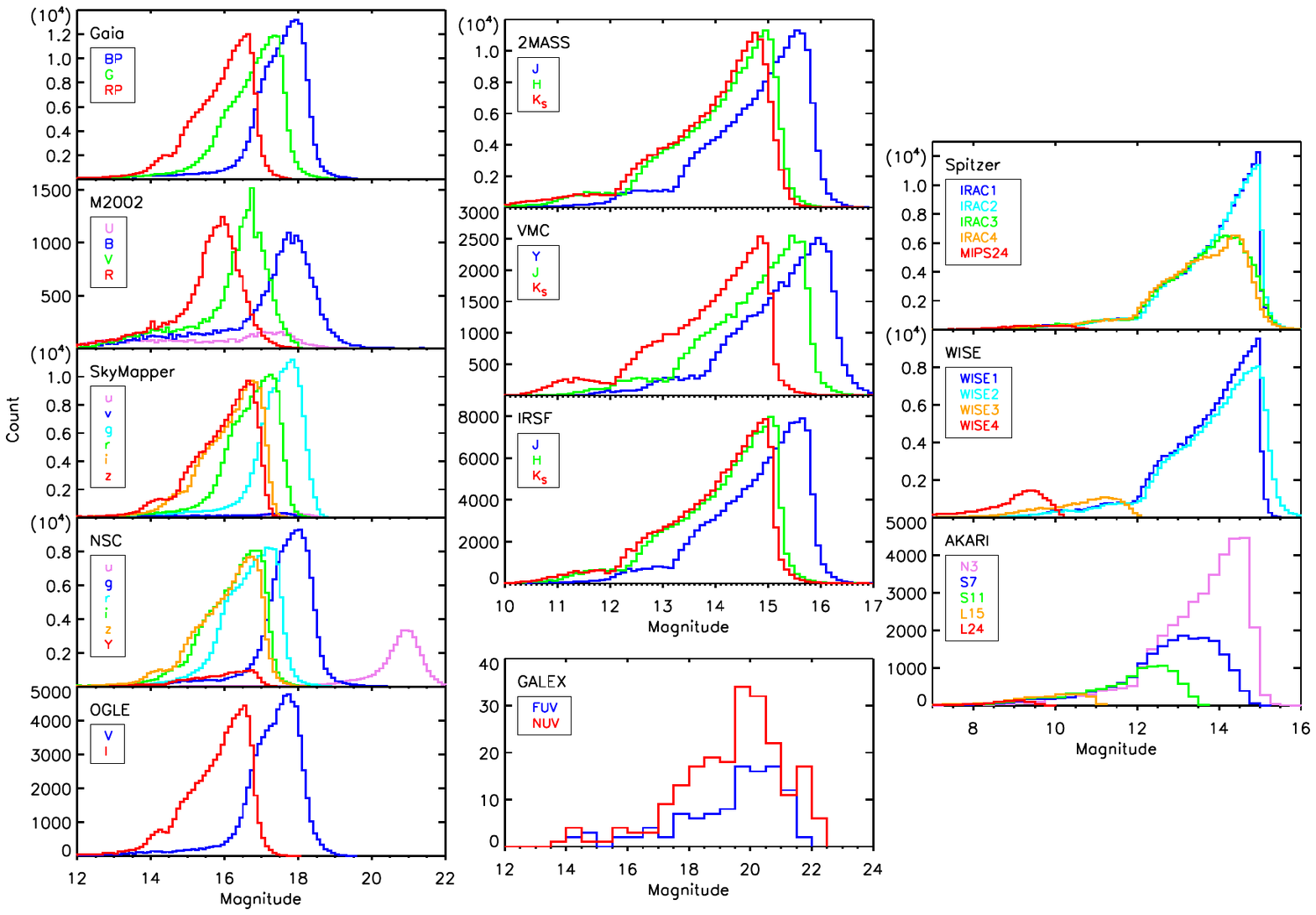}
\caption{Histograms of magnitude distribution in each dataset. The bin size of magnitude is 0.1 mag, except for the GALEX (0.5 mag) and AKARI (0.25 mag) data. For WISE3 and WISE4 bands, the histograms only show targets with $S/N\geq10$. For convenience, the HERITAGE data are not shown in the diagram. \label{sc_histo}}
\end{figure*}

As the foundation of our study, this multiwavelength source catalog with 197,004 targets ensures that we select genuine LMC members based on both astrometric and IR measurements. Figure~\ref{sc_histo} shows the histograms of magnitude distribution for each dataset (for convenience, the HERITAGE data are not shown here). Meanwhile, due to the huge amount of data required to be processed, we did not include the time-series data in our catalog. Any interested reader can retrieve the data by crossmatching our source catalog with the corresponding database. Our LMC source catalog of 197,004 targets is available in its entirety in CDS. Table~\ref{isample} shows the content of each column of the catalog. Targets without errors indicate either a 95\% confidence upper limit, or the errors are too large to be reliable (e.g., $>1.0$ mag).

\section{Identify Evolved Massive Star Candidates with the Color-Magnitude Diagrams}

To investigate evolved dusty massive stars, the first step is to identify them. However, for this paper, we decided to use modified color and magnitude cuts based on \citet{Yang2019} (see more details in Section 4 of \citealt{Yang2019}), which represented equivalent evolutionary phases (EEPs) from core helium burning to carbon burning with 7 to 40 $M_\sun$ and $A_{\rm V}=1.0$ mag (including both foreground and internal extinction), and were scaled according to the LMC distance. Notice that, these cuts may not reflect the expected changes in the position of evolutionary features with metallicity. Using modified cuts instead of evolutionary tracks is mainly because there is a problem (the EEPs do not properly fit the CMDs) with the models at LMC metallicity from Modules for Experiments in Stellar Astrophysics (MESA; \citealt{Paxton2011, Paxton2013, Paxton2015, Paxton2018}) Isochrones \& Stellar Tracks (MIST\footnote{http://waps.cfa.harvard.edu/MIST/}; \citealt{Choi2016, Dotter2016}) (Aaron Dotter, private communication). Meanwhile, the PAdova and TRieste Stellar Evolution Code (PARSEC; \citealt{Bressan2012, Tang2014, Chen2015, Chen2019}) was also investigated. Unfortunately, the critical evolutionary points of PARSEC were also inappropriate for the data, plus the rotational mixing has not yet been introduced in PARSEC (Yang Chen, private communication).

\begin{figure}
\center
\includegraphics[bb=125 365 495 690, scale=0.65]{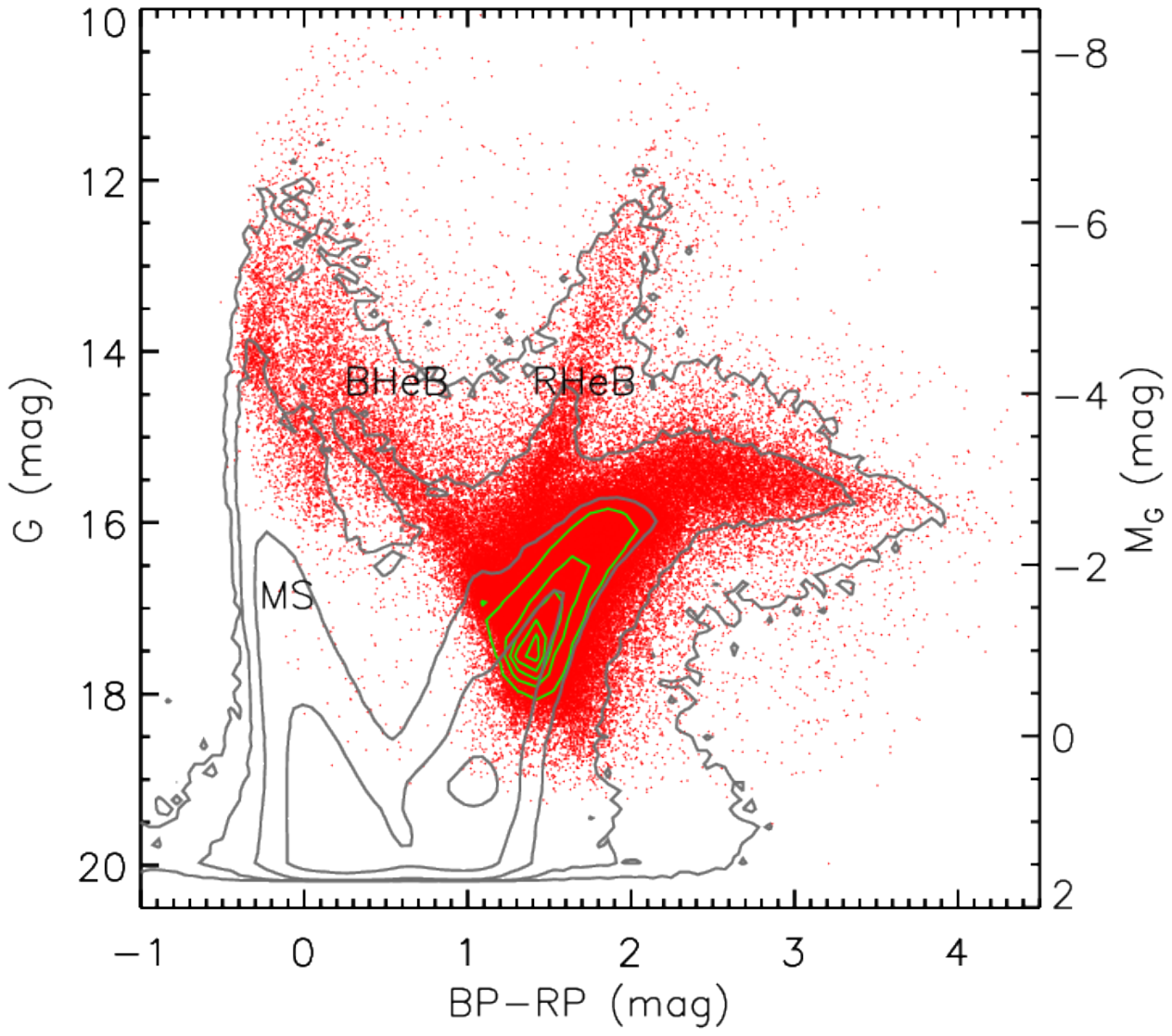}
\caption{Comparison of Gaia CMDs between our sample (red dots with green contours indicating the number density) and LMC sample from \citet{Gaia2018b} (gray contours indicating the number density). The approximate positions of MSs, BHeBs, and RHeBs for the sample of \citet{Gaia2018b} are indicated.
\label{bheb}}
\end{figure}

Firstly, Figure~\ref{bheb} shows the comparison of Gaia CMDs between our sample and LMC sample from \citet{Gaia2018b}. We indicated the approximate positions of main sequence stars (MSs), blue helium burning stars (BHeBs; stars with initial masses $\geqslant$2 $M_\sun$ and evolved off the MS with core helium burning; \citealt{DohmPalmer1997, McQuinn2011, Dalcanton2012, Schombert2015}), and red HeB stars (RHeBs) for the sample of \citet{Gaia2018b} (the BHeBs and RHeBs were referred to be BSGs and RSGs in our sample). From the diagram, it can be seen that, since our sample is selected based on the infrared criterion (IRAC1 or WISE1$\leq$15.0 mag), it presumably traces the relatively luminous, cooler evolved stars with larger MLR, like the BHeBs and RHeBs, than the MSs. Hotter stars, e.g., most of the MSs, would not be selected due to the weaker radiation at the far end of the Rayleigh-Jeans tail and/or smaller MLR than the cooler evolved stars. However, there will be also inevitably contamination of the main sequence massive stars at the blue end, which cannot be easily disentangled as shown in the diagram.

\begin{figure*}
\center
\includegraphics[bb=130 340 490 690, scale=0.6]{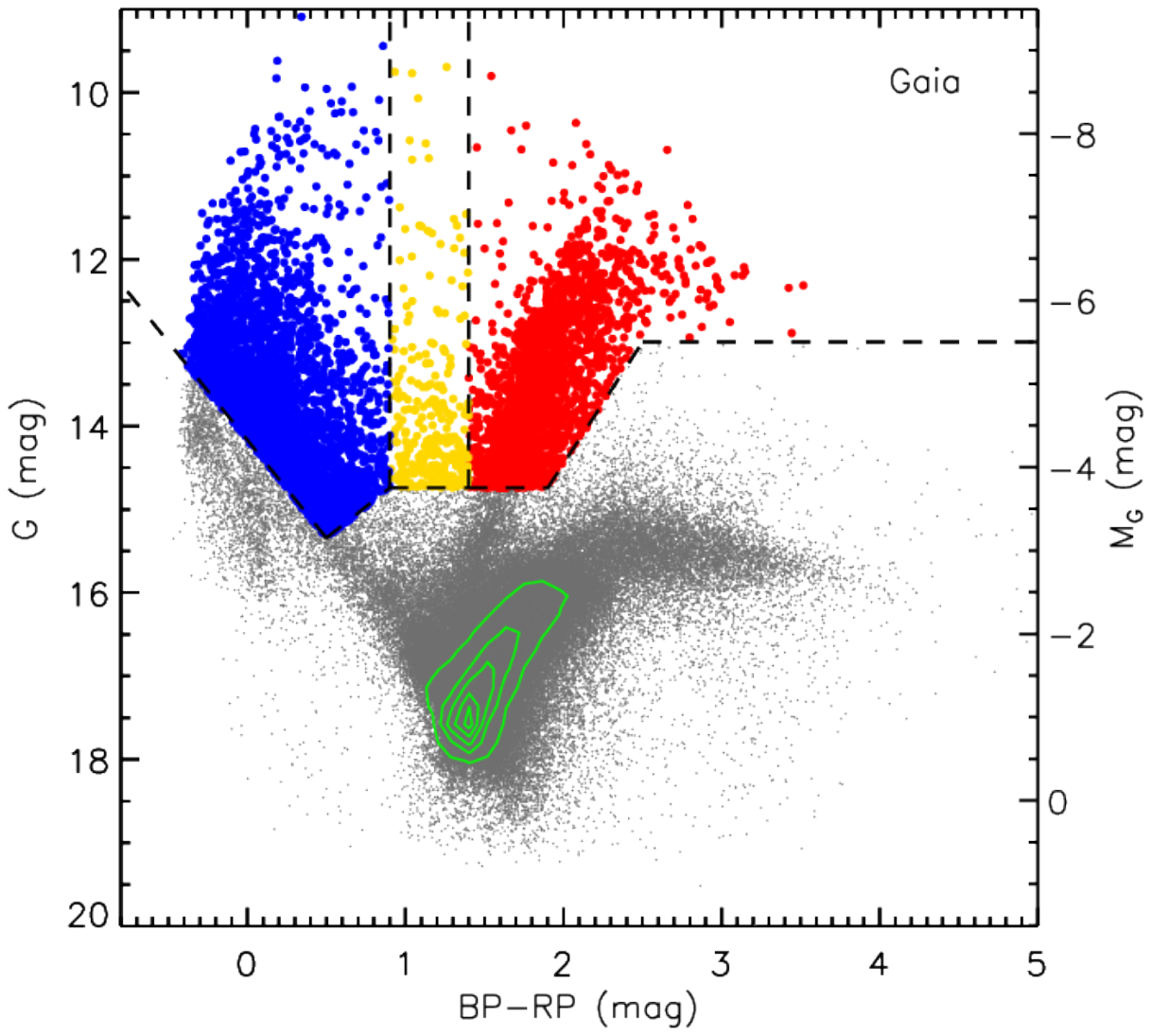}
\includegraphics[bb=130 340 490 690, scale=0.6]{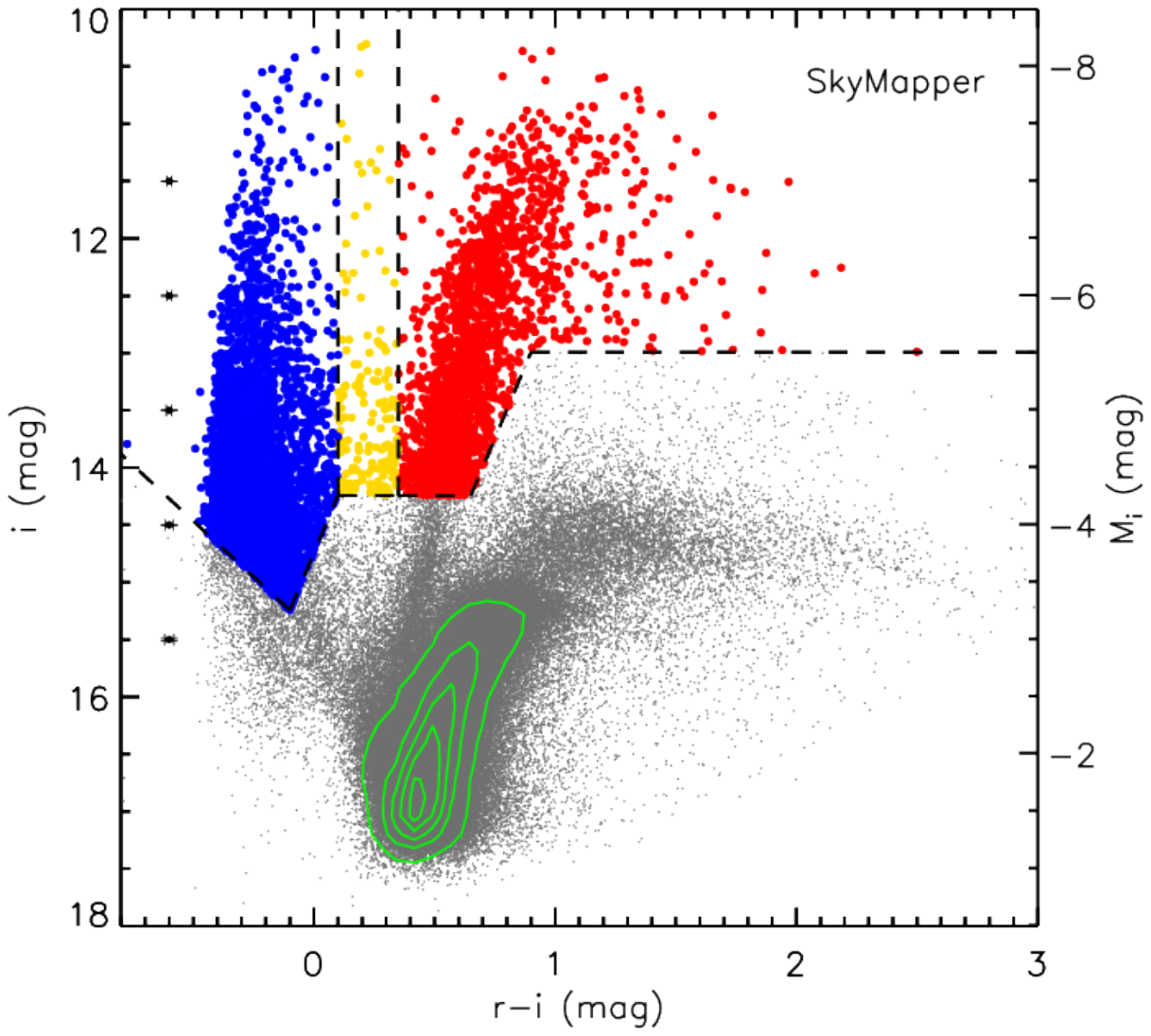}
\includegraphics[bb=130 340 490 690, scale=0.6]{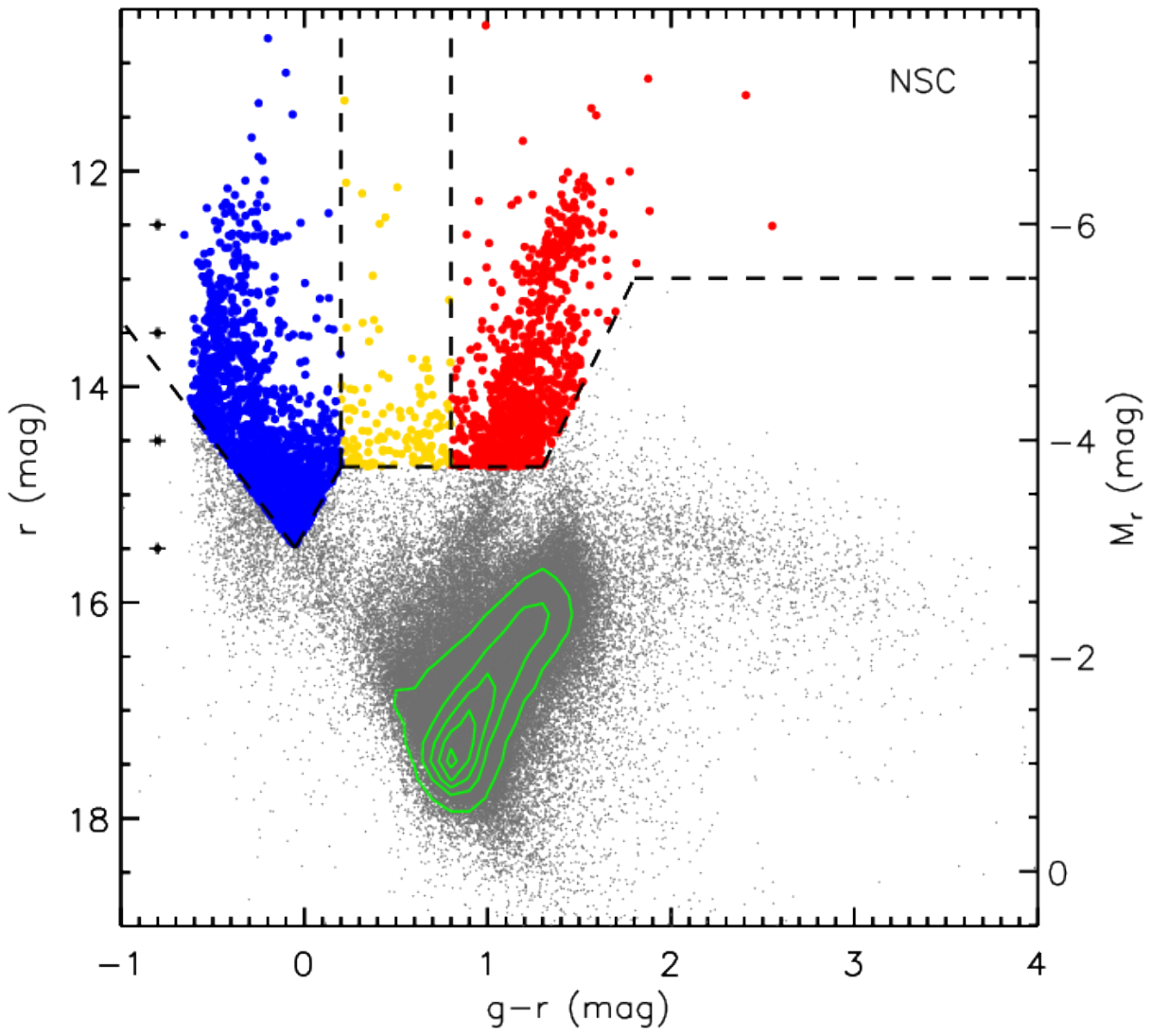}
\includegraphics[bb=130 340 490 690, scale=0.6]{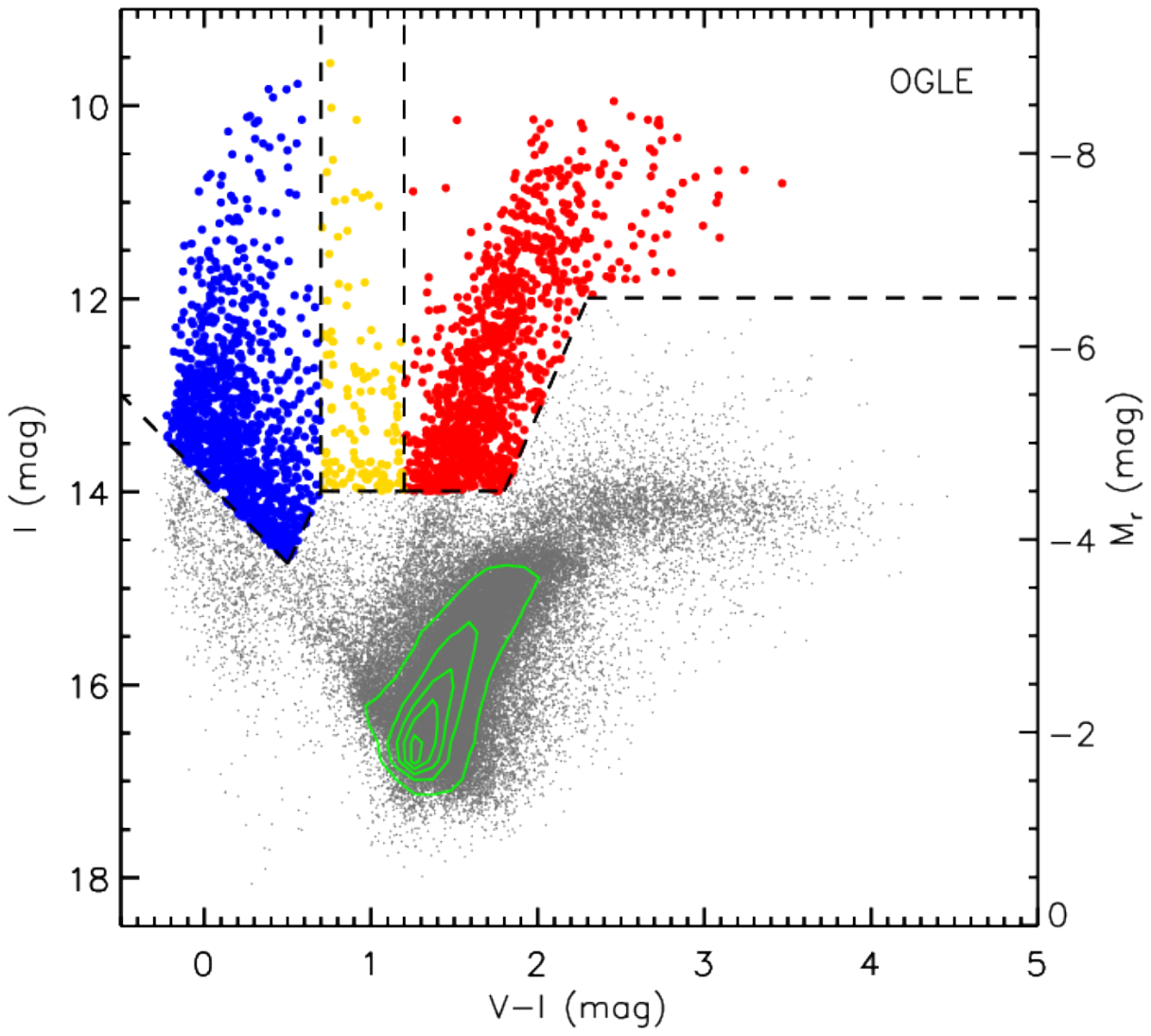}
\includegraphics[bb=130 340 490 690, scale=0.6]{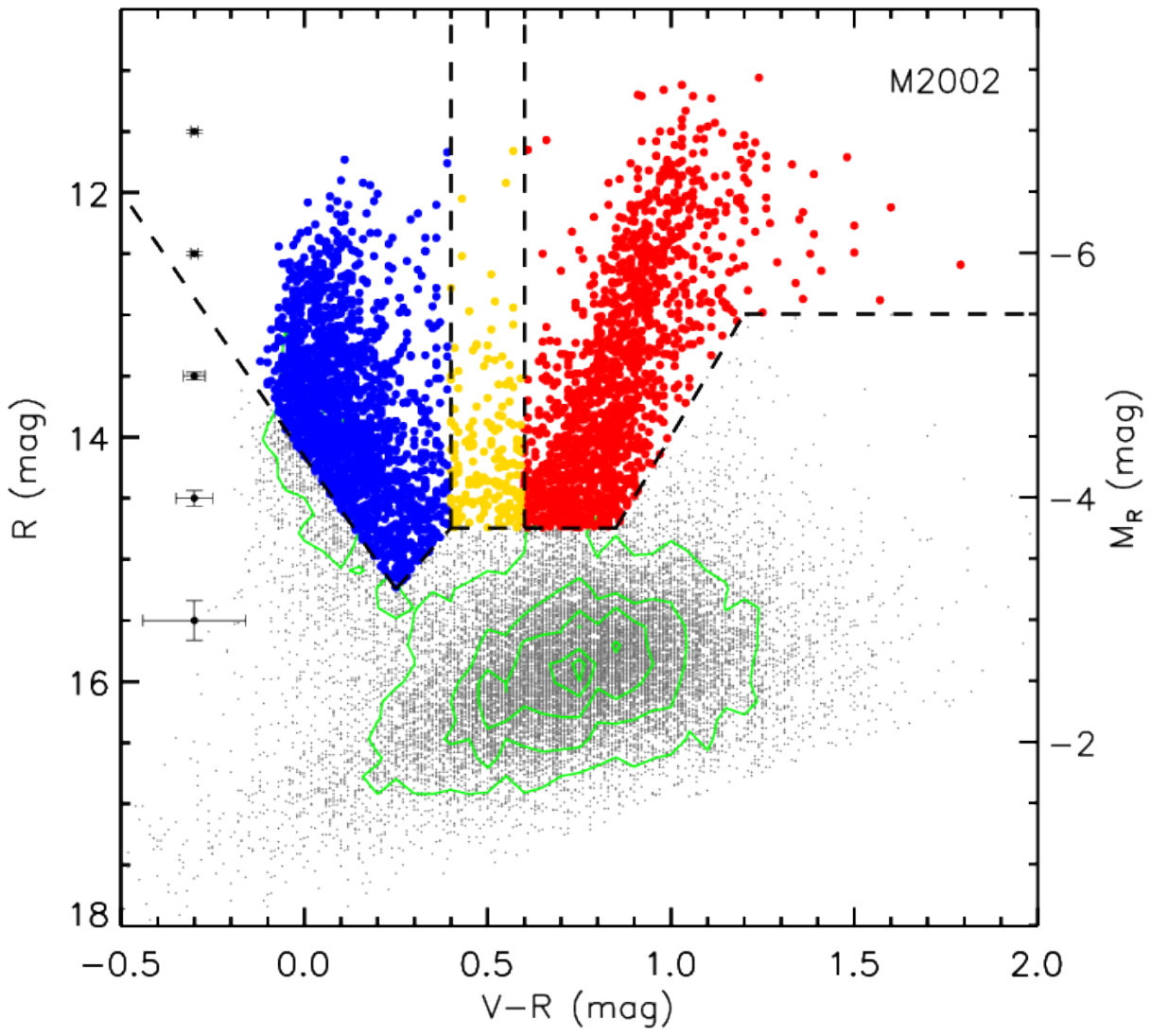}
\includegraphics[bb=130 340 490 690, scale=0.6]{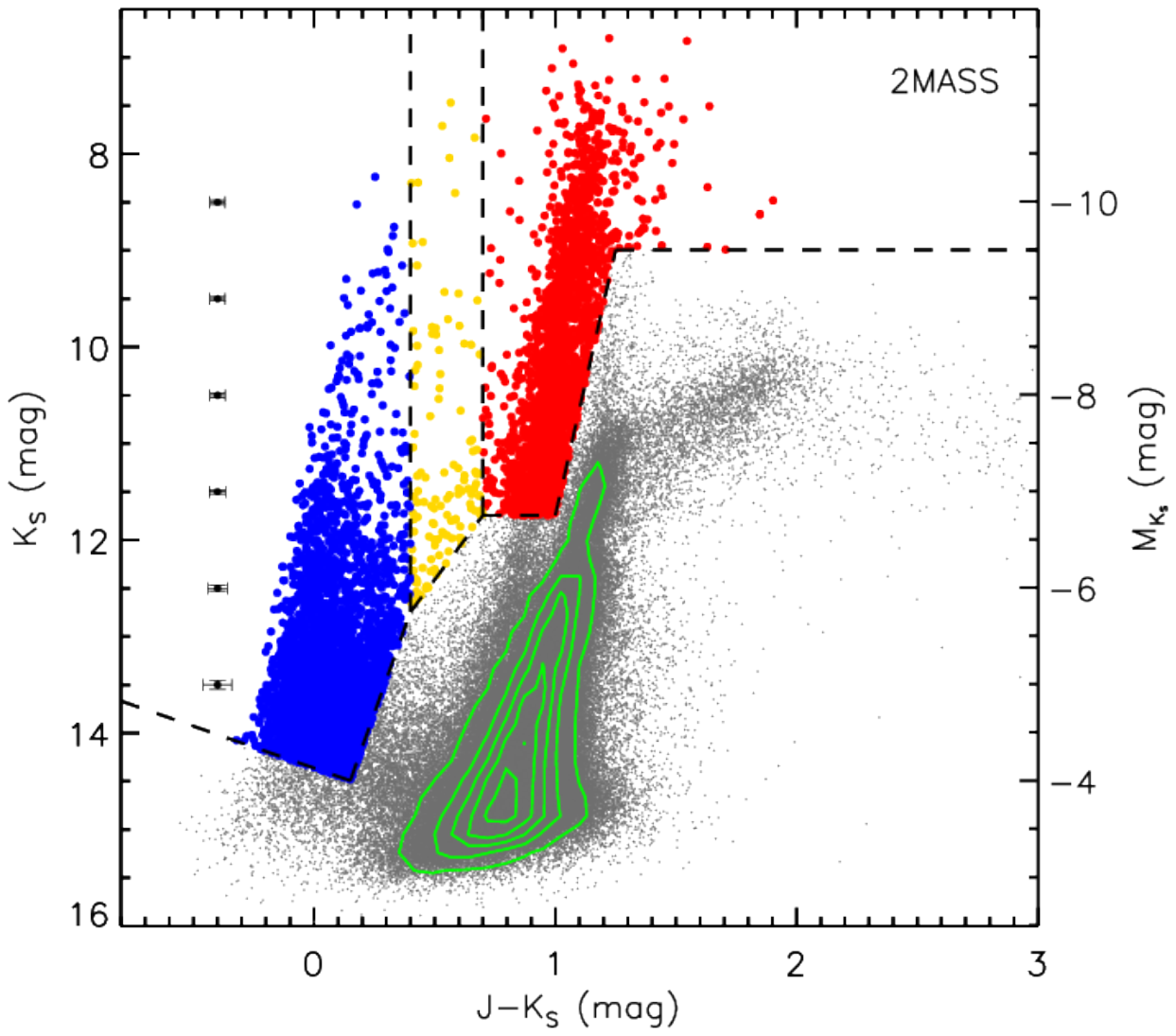}
\caption{Color-magnitude diagrams of Gaia (upper left), SkyMapper (upper right), NSC (middle left), OGLE (middle right), M2002 (bottom left), and 2MASS (bottom right) datasets. In each diagram, different regions of BSGs (blue), YSGs (yellow), and RSGs (red) are indicated by the dashed lines. The average photometric uncertainties are indicated when available. The diagrams show a clear bimodal distribution of the BSGs and RSGs candidates with few YSGs candidates lying between them.
\label{cmd_multi}}
\end{figure*}

Figure~\ref{cmd_multi} shows multiple CMDs of Gaia, SkyMapper, NSC, OGLE, M2002, and 2MASS datasets. Each type of evolved massive stars, namely BSGs (blue), YSGs (yellow), and RSGs (red), are indicated by the regions outlined by dashed lines. The criteria listed in Table~\ref{outlines} are determined based on \citet{Yang2019}, but also modified according to the morphology of each population in the CMDs. That is to say, the determination is based on a bimodal distribution of the BSG and RSG candidates with few YSG candidates lying between them. Moreover, even though the separation between RSGs and AGBs is relatively clear after the astrometric constraint (see Figure 15 of \citealt{Yang2019}), as discussed in Section 3.1 of \citet{Yang2020}, there is still a blurred boundary between RSGs and AGBs with continuity in photometry, variabilities, and even in their spectra. Currently, there is no efficient way to clearly distinguish between them. Thus, to be on the safe side, the red boundary of RSG population was set by closely resembling the limit defined by Cioni-Boyer method \citep{Cioni2006, Boyer2011, Yang2019, Yang2020}, which was bluer than the MIST boundary. Notice that, similar to the SMC \citep{Yang2019}, at the faint end of the RSG population (fainter than the limit of 7 $M_\sun$ model track scaled from the SMC models), there is a distinct branch stretching continuously towards the TRGB after the astrometric constraint, indicating a genuine lower limit of initial mass of RSGs (6$\sim$7 $M_\sun$) lower than the conventional definition ($\sim$8 $M_\sun$). However, in order to properly compare to the SMC and avoid further contamination from the red giants or AGBs, we did not set up a lower limit for the RSG population. We would also like to indicate that, up to now, there is almost no study focus on the faint end of RSG population. For example, almost all the spectroscopically confirmed RSGs in the SMC are brighter than $K_{\rm S}\approx14.0$ mag, which only represent $\sim$25\% of the sample in \citet{Yang2020}. For our LMC sample, the percentage is even surprisingly lower as $\sim$6\% with almost all the spectroscopically confirmed RSGs brighter than $K_{\rm S}\approx10.0$ mag. In that sense, we barely have real knowledge about low-mass RSGs and true difference between them and the AGBs. Moreover, some of the lowest-mass RSGs may temporarily exit the RSG phase, evolving to the left on the Hertzsprung-Russell diagram and producing ``blue loop'' as they temporarily (or permanently) return to YSGs (or even BSGs) state \citep{Maeder1994, Maeder2000, Meynet2015}. Some of them may be also related to the intermediate-luminosity optical transients (ILOTs; \citealt{Prieto2008, Bond2009, Berger2009}). Thus, low-mass RSGs connecting the evolved massive and intermediate stars are crucial for understanding star formation history and stellar evolution in nearby galaxies. More details and discussion can be found in corresponding sections of \citet{Yang2019, Yang2020}. The average photometric uncertainties are indicated when available, while targets without errors are not shown in the CMDs. 


\begin{figure*}
\center
\includegraphics[bb=60 425 550 635, scale=1.05]{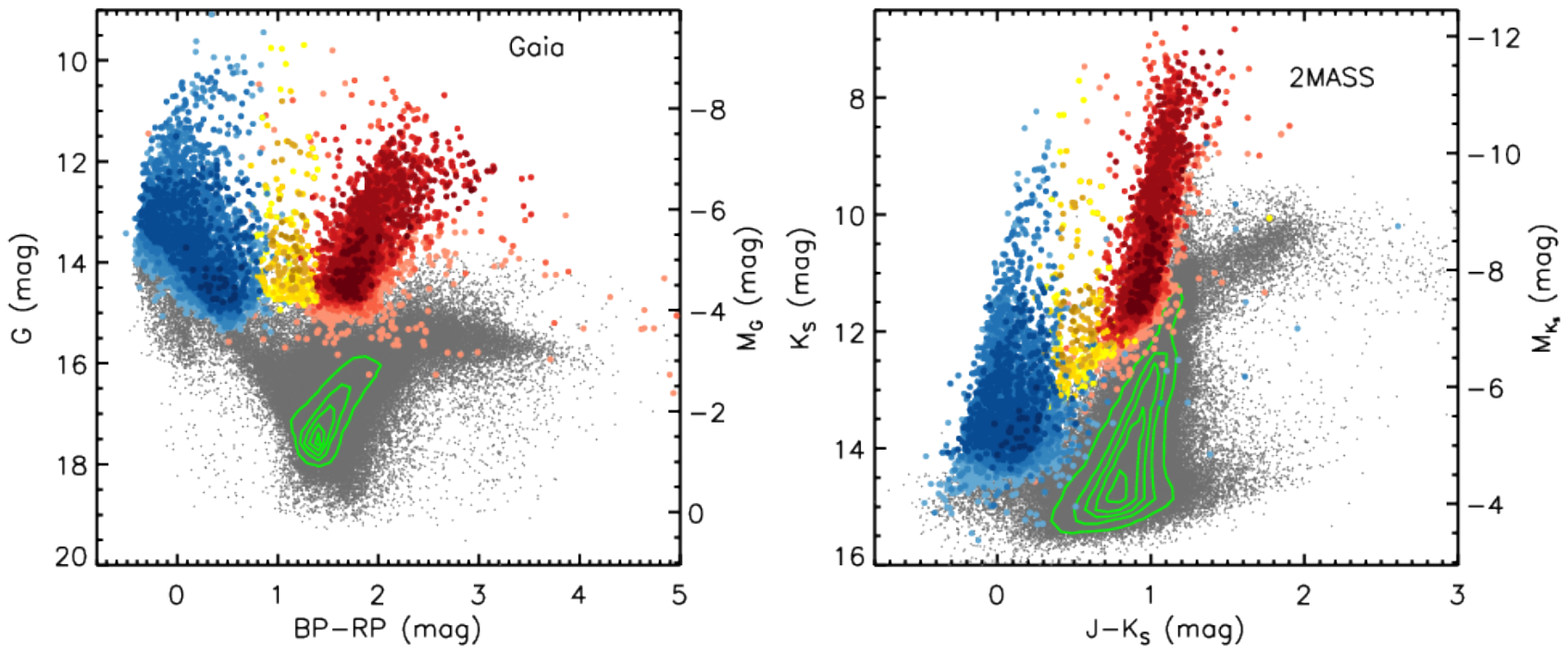}
\caption{Color-magnitude diagrams of Gaia (left) and 2MASS (right) with RSG (red), YSG (yellow), and BSG (blue) candidates overlapped, where the colors are coded from dark (Rank 0) to light (Rank 5) based on the ranks. The RSG branch extends towards fainter magnitudes with few candidates scattered in the much fainter and redder region in the optical band, which is likely caused by the circumstellar dust envelope. Green contours represent the number density. 
\label{cmd_candidates_rank}}
\end{figure*}

\begin{figure*}
\center
\includegraphics[bb=60 440 555 620, scale=1.03]{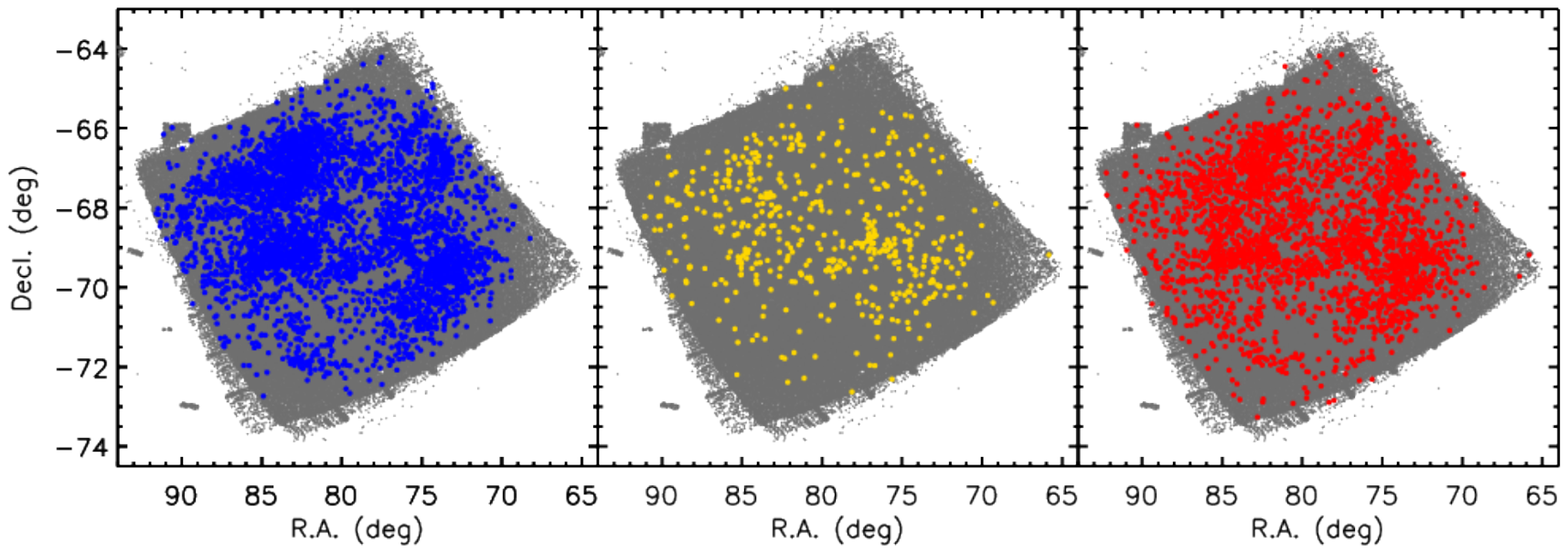}
\caption{Spatial distribution of BSG (left), YSG (middle), and RSG candidates (right). 
 \label{candi_spatial}}
\end{figure*}

The candidates of each evolved massive star populations were combined with duplications removed, resulting in 2,974 RSG, 508 YSG, and 4,786 BSG candidates in total as listed in Table~\ref{cmd_ident}. The candidates were then ranked (from 0 to 5) based on the intersection between different CMDs, where Rank 0 indicates that a target has been identified as the same type of evolved massive star in all six datasets (Gaia, SkyMapper, NSC, M2002, OGLE, and 2MASS) and so on (notice that there is no Rank 0 target for YSG candidates). The numbers of ranked candidates for each evolved massive star population are listed in Table~\ref{rank_number}. It also shows the comparison between the LMC and SMC, where the ranks are aligned based on in how many CMDs a candidate has been identified as the same type of evolved massive star. Figure~\ref{cmd_candidates_rank} shows Gaia and 2MASS CMDs for all ranked candidates color coded in red (RSGs), yellow (YSGs), and blue (BSGs) ranging from dark (Rank 0) to light (Rank 5). Detailed information about each type of evolved massive star candidates is presented in separate tables available in CDS, which have similar formats of Table~\ref{isample}. Finally, Figure~\ref{candi_spatial} shows the spatial distribution of evolved massive star candidates. 

\section{Comparison between the LMC and SMC}

\begin{figure*}
\center
\includegraphics[bb=165 360 450 725, scale=1.75]{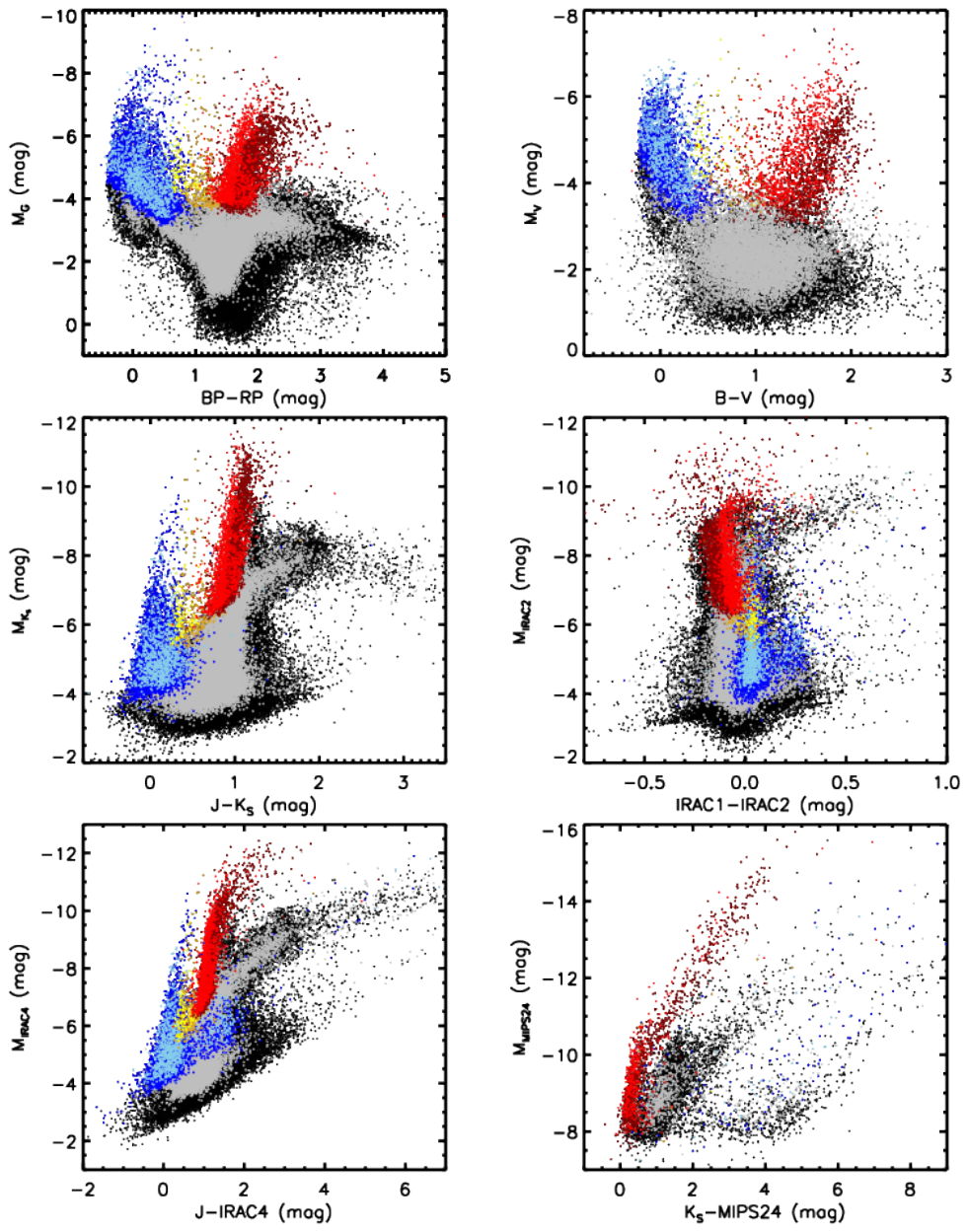}
\caption{Comparison of all targets in the LMC (black) and SMC (gray) in multiple CMDs of $M_{\rm G}$ versus $BP-RP$ (upper left), $M_{\rm V}$ versus $B-V$ (upper right), $M_{K_{\rm S}}$ versus $J-K_{\rm S}$ (middle left), $M_{\rm IRAC2}$ versus $IRAC1-IRAC2$ (middle right), $M_{\rm IRAC4}$ versus $J-IRAC4$ (bottom left), and $M_{\rm MIPS24}$ versus $K_{\rm S}-MIPS24$ (bottom right). Massive star candidates are color coded as blue (BSGs), yellow (YSGs), and red (RSGs), with dark and light colors indicating targets from the LMC and SMC, respectively. There is a prominent difference at the bright red end (RSG population) of the optical and NIR CMDs, that LMC targets show redder color than the SMC ones. Meanwhile, this trend is reversed in $M_{\rm IRAC2}$ versus $IRAC1-IRAC2$ diagram with the LMC targets showing bluer color.
\label{lmc_smc}}
\end{figure*}

As the two source catalogs for the LMC and SMC are established, we have an opportunity to compare the general properties of the two neighbor galaxies. However, it is also worth to mention that these two catalogs are only covering the main bodies of MCs, since they are based on the observation of Spitzer. In that sense, there will be more targets, as well as evolved massive star candidates, in the periphery of the MCs. A true distance modulus of 18.95$\pm0.07$ mag for the SMC was adopted \citep{Graczyk2014, Scowcroft2016}, while the adopted extinction for selecting massive star candidates is the same as the LMC ($A_{\rm V}=1.0$ mag; \citealt{Yang2019}). Figure~\ref{lmc_smc} shows the multiple CMDs ($M_{\rm G}$ versus $BP-RP$, $M_{\rm V}$ versus $B-V$, $M_{K_{\rm S}}$ versus $J-K_{\rm S}$, $M_{\rm IRAC2}$ versus $IRAC1-IRAC2$, $M_{\rm IRAC4}$ versus $J-IRAC4$, $M_{\rm MIPS24}$ versus $K_{\rm S}-MIPS24$) from optical to MIR for the source catalogs of the LMC (black) and SMC (gray). Massive star candidates are also overplotted and color coded as blue (BSGs), yellow (YSGs), and red (RSGs), with dark and light colors indicating targets from the LMC and SMC, respectively. It can be seen from the diagram that, the most distinct difference between these two galaxies appears at the bright red end of the optical and NIR CMDs, where the cool evolved stars are located. The LMC targets, including RSGs, AGB, and RGs, are redder than the SMC ones, which is likely due to different metallicities or star formation histories (SFHs) between the two galaxies as discussed below. 

\begin{figure*}
\center
\includegraphics[bb=55 380 560 700, scale=1.]{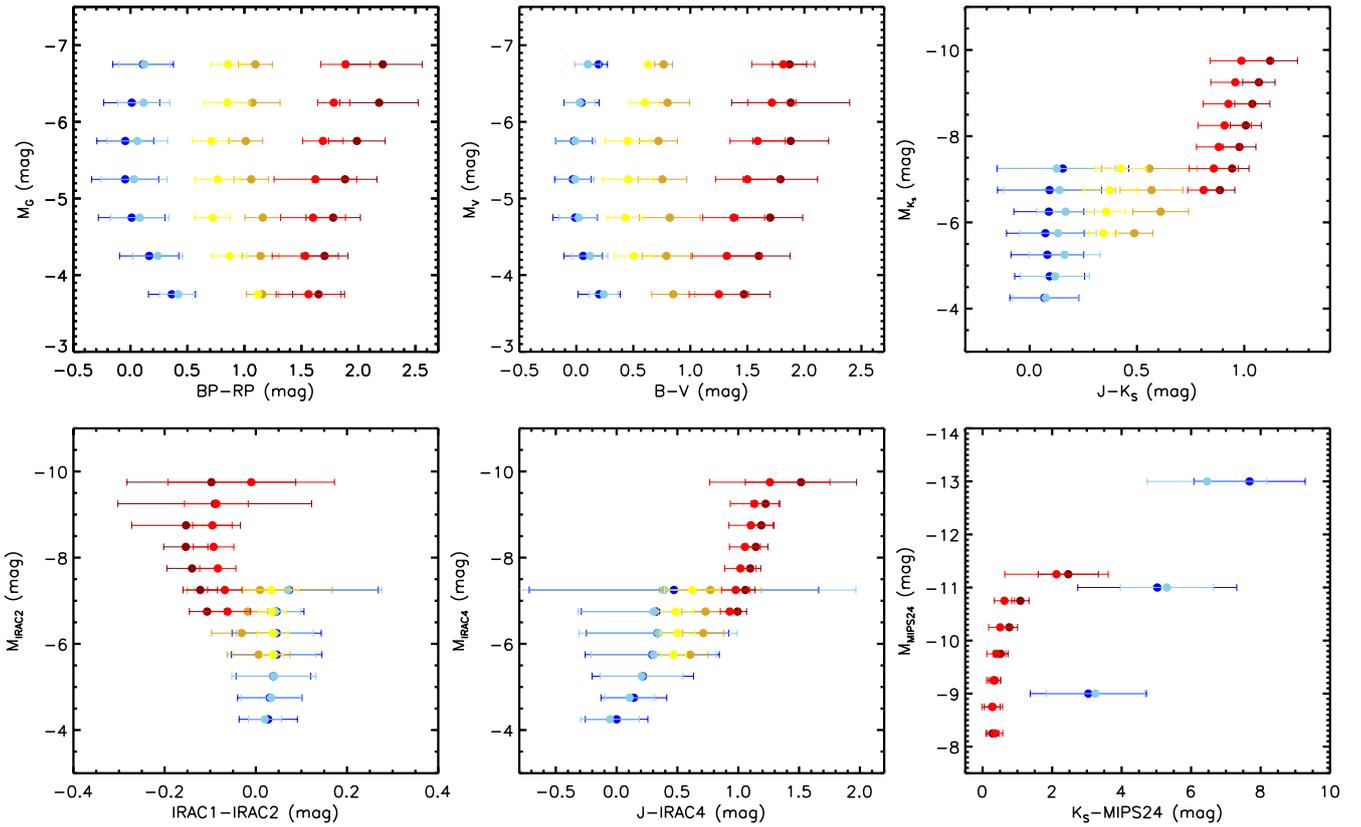}
\caption{Same as Figure~\ref{lmc_smc}, but only for massive star candidates selected in at least two CMDs (Rank 0 to 4 in the LMC and Rank 0 to 3 in the SMC). For each diagram, targets are divided into equal magnitude bins and the values of median (solid circles) and SD (errors) of color in each bin are calculated.
\label{lmc_smc_color}}
\end{figure*}

Moreover, we also quantitatively compared the colors of massive star populations between these two galaxies as shown in Figure~\ref{lmc_smc_color}. To be on the safe side, the comparison was only based on candidates selected in at least two CMDs (Rank 0 to 4 in the LMC and Rank 0 to 3 in the SMC). Targets were divided into equal magnitude bins (except the very bright end), for which the values of median and standard deviation (SD) of color in each bin were calculated as summarized in Table~\ref{ctable}. It can be seen that there is almost no difference for the median values of BSG candidates between the LMC and SMC. The difference starts to emerge from YSG to RSG candidates, as LMC targets are getting redder, which may be due to combined effect of the several factors. One is that the average spectral type of RSGs moves towards earlier types at lower metallicities, due to the lower opacity leads to higher effective temperature measured at deeper layer of the stellar photosphere, and/or the metallicity-dependent Hayashi limit shifts to warmer temperatures at lower metallicities \citep{Hayashi1961, Elias1985, Massey2003, Levesque2006, Levesque2012, Dorda2016}. The other is that there is a positive relation between mass-loss rate (MLR) and metallicity \citep{vanLoon2005, Mauron2011}, as a metallicity-scaling factor of $(Z/Z_\sun)^{0.7}$ may be applied to the \citet{deJager1988} prescription which results in about 1.6 times higher MLR in the LMC than in the SMC. However, this difference is reversed in $IRAC1-IRAC2$ color (LMC targets are bluer than the SMC ones, the same as $WISE1-WISE2$), which may indicate that there is a negative relation between CO absorption around 4.6 $\mu$m~and the metallicity for RSGs. Moreover, the interaction between the MW, the LMC, and the SMC, which triggers multiple peaks of star formation in the past few hundreds million years \citep{Yoshizawa2003, Harris2009, Indu2011}, may have different effect on the two galaxies. For example, \citet{Bitsakis2017, Bitsakis2018} studied the distribution and ages of star clusters in the LMC and SMC and found that the star clusters at the central regions were younger in the SMC than in the LMC. So that if SMC RSGs are younger then perhaps this may indicate that they have lost less mass (assumed the MLR is the same), so less material is around them (thus they appear less redder than the LMC ones).

\begin{figure}
\center
\includegraphics[bb=120 365 455 690, scale=0.65]{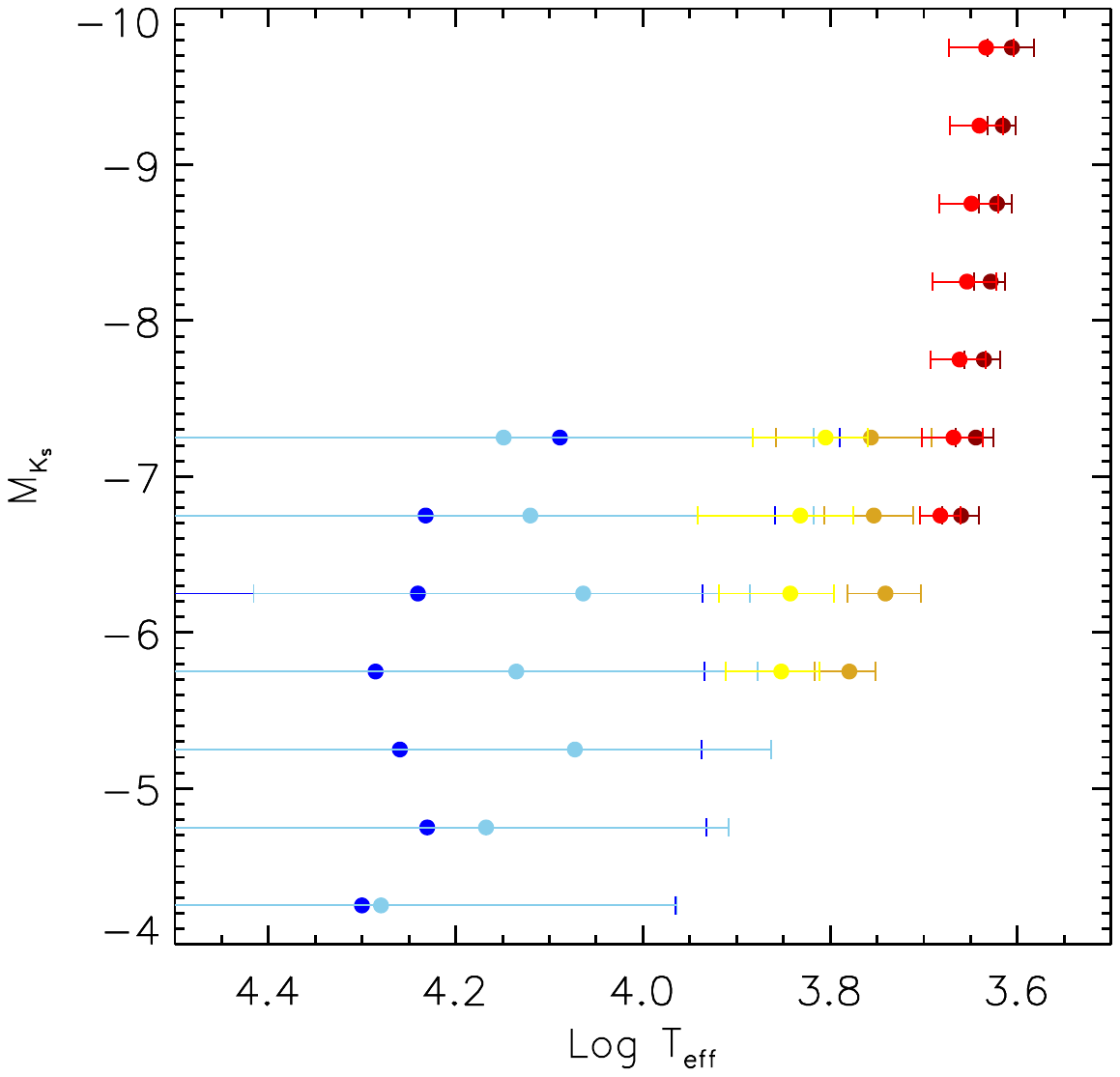}
\caption{$T_{\rm eff}$ of massive star populations derived from reddening-free color of $(J-K_{\rm S})_0$. The $T_{\rm eff}$ ranges are $3500<T_{\rm eff}<5000$ K for RSG population, $5000<T_{\rm eff}<8000$ K for YSG population, and $T_{\rm eff}>8000$ K for BSG population. Errors indicate the SD of the $T_{\rm eff}$.
\label{lmc_smc_teff}}
\end{figure}

We also converted the reddening-free color of $(J-K_{\rm S})_0$ (by adopting $A_{\rm V}=1.0$ mag, $A_{K_{\rm S}}/A_{\rm V}\approx0.1$, and $A_{\rm J}/A_{K_{\rm S}}=3.12$; \citealt{Wang2019}) of the massive star populations into the effective temperature ($T_{\rm eff}$) by using Equation (3) of \citet{Yang2020} as shown in Figure~\ref{lmc_smc_teff}. As expected, the RSG population is located in a narrow range of about $3500<T_{\rm eff}<5000$ K, YSG population is around $5000<T_{\rm eff}<8000$~K, and BSG population is much more broad as $T_{\rm eff}>8000$ K. The difference of $T_{\rm eff}$ between the LMC and SMC for each population also increases as $\sim$300 K for RSG, $\sim$1500 K for YSG, and more than $\sim$5000 K for BSG population, respectively. However, we also would like to indicate that the estimation of $T_{\rm eff}$ for the hotter stars (e.g., BSGs) may suffer larger uncertainties compared to cooler stars (e.g., RSGs) due to the weak emission at the NIR wavelength as discussed in previous section.

In addition, we would like to emphasize one important thing as mentioned in previous section and also in \citet{Yang2019}. Since our sample was selected based on the infrared criteria and constrained by multiple factors (deblending, astrometry, model limitation), hotter stars would be more incomplete than the cooler stars. Thus, the BSG to RSG ratio (B/R ratio) should not be simply assumed as $\sim$1.5:1 in the LMC or $\sim$1:1 in the SMC (e.g., see Table~\ref{rank_number}), since both observation and theoretical prediction suggest more BSGs than RSGs (e.g., B/R ratio $\sim$4 or more) in the MCs \citep{Meylan1982, Humphreys1984, Guo2002}.

\section{Summary}

We present a clean, magnitude-limited (IRAC1 or WISE1$\leq$15.0 mag) multiwavelength source catalog for the LMC. The catalog contains 197,004 targets with data in 52 different bands including 2 UV, 21 optical and 29 infrared bands, retrieved from SEIP, VMC, IRSF, AKARI, HERITAGE, Gaia, SkyMapper, NSC, OGLE, M2002, and GALEX datasets, ranging from ultraviolet to far-infrared.  Additional information about radial velocities and spectral/photometric classifications were collected from the literature. 

The same method of \citet{Yang2019} was applied to establish the catalog, that it was built upon crossmatching ($1''$) and deblending ($3''$) between SEIP source list and Gaia DR2 photometric data, with strict constraints on the PMs and parallaxes from Gaia DR2 to remove the foreground contamination. It was estimated that about 99.5\% of the targets in our catalog were most likely to be genuine members of the LMC.

We compared our sample with sample from \citet{Gaia2018b}, indicating that the bright end of our sample was mostly comprised of BHeB and RHeB with inevitable contamination of MSs at the blue end. We applied modified magnitude and color cuts based on our previous works \citep{Yang2019, Yang2020} on six different CMDs, and identified 2,974 RSG, 508 YSG, and 4,786 BSG candidates in the LMC. The candidates were ranked according to the intersections among all available CMDs, for which those with the highest number (six) given the highest rank. 

The comparison between the source catalogs of the LMC and SMC indicates that, the most distinct difference between these two galaxies appears at the bright red end of the optical and NIR CMDs, where the cool evolved stars are located. The LMC targets are redder than the SMC ones, which is likely due to the effect of metallicity and SFH. Meanwhile, the quantitative comparison of colors of massive star candidates in equal absolute magnitude bins shows that, there is basically no difference for the BSG candidates. However, there is an obvious discrepancy for the RSG candidates as LMC targets are redder than the SMC ones, which may be due to the combined effect of metallicity on both spectral type and MLR, and also the age effect. $T_{\rm eff}$ of massive star populations were also derived from reddening-free color of $(J-K_{\rm S})_0$. The $T_{\rm eff}$ ranges are $3500<T_{\rm eff}<5000$ K for RSG population, $5000<T_{\rm eff}<8000$ K for YSG population, and $T_{\rm eff}>8000$ K for BSG population, with larger uncertainties towards the hotter stars.

\begin{acknowledgements}

This study has received funding from the European Research Council (ERC) under the European Union's Horizon 2020 research and innovation programme (grant agreement number 772086). B.W.J and J.G. gratefully acknowledge support from the National Natural Science Foundation of China (Grant No.11533002 and U1631104). We thank Dr. Yang Chen for providing valuable comments on the stellar evolution models of PARSEC.

This publication makes use of data products from the Two Micron All Sky Survey, which is a joint project of the University of Massachusetts and the Infrared Processing and Analysis Center/California Institute of Technology, funded by the National Aeronautics and Space Administration and the National Science Foundation. This work is based in part on observations made with the Spitzer Space Telescope, which is operated by the Jet Propulsion Laboratory, California Institute of Technology under a contract with NASA. This publication makes use of data products from the Wide-field Infrared Survey Explorer, which is a joint project of the University of California, Los Angeles, and the Jet Propulsion Laboratory/California Institute of Technology. It is funded by the National Aeronautics and Space Administration. This publication makes use of data products from the Near-Earth Object Wide-field Infrared Survey Explorer (NEOWISE), which is a project of the Jet Propulsion Laboratory/California Institute of Technology. NEOWISE is funded by the National Aeronautics and Space Administration. This research has made use of the NASA/IPAC Infrared Science Archive, which is operated by the Jet Propulsion Laboratory, California Institute of Technology, under contract with the National Aeronautics and Space Administration.

This work has made use of data from the European Space Agency (ESA) mission {\it Gaia} (\url{https://www.cosmos.esa.int/gaia}), processed by the {\it Gaia} Data Processing and Analysis Consortium (DPAC, \url{https://www.cosmos.esa.int/web/gaia/dpac/consortium}). Funding for the DPAC has been provided by national institutions, in particular the institutions participating in the {\it Gaia} Multilateral Agreement.

This research uses services or data provided by the NOAO Data Lab. NOAO is operated by the Association of Universities for Research in Astronomy (AURA), Inc. under a cooperative agreement with the National Science Foundation.

The national facility capability for SkyMapper has been funded through ARC LIEF grant LE130100104 from the Australian Research Council, awarded to the University of Sydney, the Australian National University, Swinburne University of Technology, the University of Queensland, the University of Western Australia, the University of Melbourne, Curtin University of Technology, Monash University and the Australian Astronomical Observatory. SkyMapper is owned and operated by The Australian National University's Research School of Astronomy and Astrophysics. The survey data were processed and provided by the SkyMapper Team at ANU. The SkyMapper node of the All-Sky Virtual Observatory (ASVO) is hosted at the National Computational Infrastructure (NCI). Development and support the SkyMapper node of the ASVO has been funded in part by Astronomy Australia Limited (AAL) and the Australian Government through the Commonwealth's Education Investment Fund (EIF) and National Collaborative Research Infrastructure Strategy (NCRIS), particularly the National eResearch Collaboration Tools and Resources (NeCTAR) and the Australian National Data Service Projects (ANDS).

This research has made use of the SIMBAD database and VizieR catalog access tool, operated at CDS, Strasbourg, France, and the Tool for OPerations on Catalogues And Tables (TOPCAT; \citealt{Taylor2005}).

Based on data products from observations made with ESO Telescopes at the La Silla or Paranal Observatories under ESO programme ID 179.B-2003.

\end{acknowledgements}

\begin{table*}
\footnotesize
\caption{LMC Source Catalog Contents}
\label{isample}
\centering
\begin{tabular}{lll}
\toprule\toprule
Column  & Name                                		& Description \\
1       & ID                                  		& Index \\
2       & R.A.(J2000)                         		& Right Ascension, J2000 (deg) \\
3       & Decl.(J2000)                        		& Declination, J2000 (deg) \\
4-9     & 2MASS\_J, ..., e\_2MASS\_$K_{\rm S}$      		& 2MASS photometries with uncertainties (mag) \\
10-19   & IRAC1, ..., e\_MIPS24               		& Spitzer photometries with uncertainties (mag) \\
20-27   & WISE1, ..., e\_WISE4                		& WISE photometries with uncertainties (mag) \\
28-33   & Gaia\_parallax, ..., e\_Gaia\_pmdec 		& Gaia DR2 astrometric solutions with uncertainties (mas; mas/yr) \\
34-36   & Gaia\_G, ..., Gaia\_RP         		    & Gaia DR2 mean magnitudes (mag) \\
37-38   & Gaia\_RV, e\_Gaia\_RV               		& Gaia DR2 radial velocity with uncertainty (km/s) \\
39-44   & VMC\_Y, ..., e\_VMC\_$K_{\rm S}$          		& VMC photometries with uncertainties (mag) \\
45-50   & IRSF\_J, ..., e\_IRSF\_$K_{\rm S}$              & IRSF photometries with uncertainties (mag) \\
51-60   & AKARI\_N3, ..., e\_AKARI\_L24             & AKARI photometries with uncertainties (mag) \\
61-72   & HERITAGE\_f70, ..., e\_HERITAGE\_f500     & HERITAGE fluxes with uncertainties (mJy) \\
73-84   & SkyMapper\_u, ..., e\_SkyMapper\_z        & SkyMapper photometries with uncertainties (mag) \\
85-96   & NSC\_u, ..., e\_NSC\_Y 		            & NSC photometries with uncertainties (mag) \\
97-104  & M2002\_V, ..., e\_M2002\_V\_R 		    & M2002 photometries with uncertainties (mag) \\
105-106 & OGLE\_V, OGLE\_I                          & OGLE photometries (mag) \\
107-110 & GALEX\_FUV, ..., e\_GALEX\_NUV 		    & GALEX photometries with uncertainties (mag) \\
111     & W2008\_YSO                                & YSO classifications from \citet{Whitney2008} \\
112     & B2011\_SAGE\_Class                        & IR color classifications from \citet{Boyer2011} \\
113     & J2017\_IRS\_Class                         & MIR spectral classifications from \citet{Jones2017} \\
114-115 & M2003\_RV, M2003\_MK\_Class               & Optical spectral classifications from \citet{Massey2003} (km/s) \\
116     & B2009\_MK\_Class                          & Optical spectral classifications from \citet{Bonanos2009} \\
117-118 & N2012\_RV, N2012\_MK\_Class               & Optical spectral classifications from \citet{Neugent2012} (km/s) \\
119-121 & E2015\_RV, ..., E2015\_MK\_Class          & Optical spectral classifications from \citet{Evans2015} (km/s) \\
122-124 & GF2015\_MK\_SpVar, ..., GF2015\_MK\_Class & Optical spectral classifications from \citet{Gonzalez2015} (km/s) \\
125-128 & Simbad\_RV, ..., Simbad\_Other\_Types     & Simbad classifications (km/s) \\
\midrule  
\end{tabular}
\tablefoot{
This table is available in its entirety in CDS. \\
}
\end{table*}

\begin{table*}
\caption{Evolved Massive Star Candidate Selection Criteria} 
\label{outlines}
\begin{tabular}{ccc}
\toprule\toprule
Group & Color Criteria & Magnitude Criteria \\
\midrule
BSG$_{Gaia}$ & $(BP-RP)<0.5$          & $G\leq2.350\times(BP-RP)+14.168$  \\
             & $0.5\leq(BP-RP)<0.9$   & $G\leq-1.500\times(BP-RP)+16.093$ \\
YSG$_{Gaia}$ & $0.9\leq(BP-RP)<1.4$   & $G\leq14.743$                     \\
RSG$_{Gaia}$ & $1.4\leq(BP-RP)<1.9$   & $G\leq14.743$  \\
             & $1.9\leq(BP-RP)<2.5$   & $G\leq-2.917\times(BP-RP)+20.285$ \\
             & $2.5\leq(BP-RP)$       & $G\leq12.993$                     \\
\midrule
BSG$_{SkyMapper}$ & $(r-i)<-0.1$         & $i\leq1.944\times(r-i)+15.437$  \\
                  & $-0.1\leq(r-i)<0.1$ & $i\leq-5.000\times(r-i)+14.743$ \\
YSG$_{SkyMapper}$ & $0.1\leq(r-i)<0.35$   & $i\leq14.243$                   \\
RSG$_{SkyMapper}$ & $0.35\leq(r-i)<0.65$  & $i\leq14.243$                   \\
                  & $0.65\leq(r-i)<0.9$  & $i\leq-5.000\times(r-i)+17.493$ \\
                  & $0.9\leq(r-i)$       & $i\leq12.993$                   \\
\midrule
BSG$_{NSC}$ & $(g-r)<0.05$        & $r\leq2.222\times(g-r)+15.604$  \\
            & $0.05\leq(g-r)<0.2$ & $r\leq-3.000\times(g-r)+15.343$ \\
YSG$_{NSC}$ & $0.2\leq(g-r)<0.8$  & $r\leq14.743$                   \\
RSG$_{NSC}$ & $0.8\leq(g-r)<1.3$  & $r\leq14.743$  \\
            & $1.3\leq(g-r)<1.8$  & $r\leq-3.500\times(g-r)+19.293$ \\
            & $1.8\leq(g-r)$      & $r\leq12.993$                   \\
\midrule
BSG$_{OGLE}$  & $(V-I)<0.5$         & $R\leq1.750\times(V-I)+13.868$  \\
              & $0.5\leq(V-I)<0.7$  & $R\leq-3.750\times(V-I)+16.618$ \\
YSG$_{OGLE}$  & $0.7\leq(V-I)<1.2$  & $R\leq13.993$                   \\
RSG$_{OGLE}$  & $1.2\leq(V-I)<1.8$  & $R\leq13.993$  \\
              & $1.8\leq(V-I)<2.3$  & $R\leq-4.000\times(V-I)+21.193$  \\
              & $2.3\leq(V-I)$      & $R\leq11.993$                   \\            
\midrule
BSG$_{M2002}$ & $(V-R)<0.25$        & $R\leq4.333\times(V-R)+14.160$  \\
              & $0.25\leq(V-R)<0.4$ & $R\leq-3.333\times(V-R)+16.076$ \\
YSG$_{M2002}$ & $0.4\leq(V-R)<0.6$  & $R\leq14.743$                   \\
RSG$_{M2002}$ & $0.6\leq(V-R)<0.85$ & $R\leq14.743$  \\
              & $0.85\leq(V-R)<1.2$ & $R\leq-5.000\times(V-R)+18.993$  \\
              & $1.2\leq(V-R)$      & $R\leq12.993$                   \\
\midrule
BSG$_{2MASS}$ & $(J-K_{\rm S})<0.15$         & $K_{\rm S}\leq0.870\times(J-K_{\rm S})+14.363$  \\
              & $0.15\leq(J-K_{\rm S})<0.4$  & $K_{\rm S}\leq-7.000\times(J-K_{\rm S})+15.543$  \\
YSG$_{2MASS}$ & $0.4\leq(J-K_{\rm S})<0.7$   & $K_{\rm S}\leq-3.333\times(J-K_{\rm S})+14.076$  \\
RSG$_{2MASS}$ & $0.7\leq(J-K_{\rm S})<1.0$   & $K_{\rm S}\leq11.743$ \\
              & $1.0\leq(J-K_{\rm S})<1.25$  & $K_{\rm S}\leq-11.000\times(J-K_{\rm S})+22.743$ \\
              & $1.25\leq(J-K_{\rm S})$      & $K_{\rm S}\leq8.993$                      \\
\midrule
\end{tabular}
\end{table*}

\begin{table*}
\caption{Numbers of Identified Evolved Massive Star Candidates}
\label{cmd_ident}
\begin{tabular}{ccccc}
\toprule\toprule
Dataset   & RSGs & YSGs & BSGs \\
\midrule
Gaia      & 2285 & 297 & 3368 \\
SkyMapper & 1994 & 152 & 3325 \\
NSC       & 915  & 135 & 1609 \\
M2002     & 1325 & 176 & 1721 \\
OGLE      & 1094 & 124 & 1024 \\
2MASS     & 2266 & 123 & 3182 \\
\midrule
Total (cleaned) & 2974 & 508 & 4786 \\ 
\midrule
\end{tabular}
\end{table*}

\begin{table*}
\caption{Numbers of Ranked Evolved Massive Star Candidates in the LMC and SMC}
\label{rank_number}
\begin{tabular}{ccccccc}
\toprule\toprule
Ranks  & RSG$_{\rm LMC}$ & YSG$_{\rm LMC}$ & BSG$_{\rm LMC}$ & RSG$_{\rm SMC}$ & YSG$_{\rm SMC}$ & BSG$_{\rm SMC}$ \\
\midrule
Rank 0 &  90 &   0 &   24 &     &    &     \\
Rank 1 & 574 &  14 &  631 & 352 & 6  & 239 \\
Rank 2 & 848 &  50 & 1187 & 421 & 17 & 362 \\
Rank 3 & 651 &  87 & 1177 & 229 & 44 & 315 \\
Rank 4 & 313 & 119 &  884 & 105 & 51 & 188 \\
Rank 5 & 498 & 238 &  883 & 206 & 99 & 265 \\
\midrule
Total  & 2974 & 508 & 4786 & 1313 & 217 & 1369 \\ 
\midrule
\end{tabular}
\tablefoot{
The ranks between the LMC and SMC are aligned based on in how many CMDs a candidate has been identified as the same type of evolved massive star.
}
\end{table*}

\longtab{
\footnotesize
\begin{longtable}{ccccccccc}
\caption{Color distribution of evolved massive star candidates between the LMC and SMC \label{ctable}} \\
\toprule\toprule
Type    & Magnitude range                & Color             & $N_{\rm LMC}$ & $Median_{\rm LMC}$ & $SD_{\rm LMC}$ & $N_{\rm SMC}$ & $Median_{\rm SMC}$ & $SD_{\rm SMC}$ \\
        & (mag)                          &                   &         & (mag)        & (mag)        &         & (mag)        & (mag)        \\
\midrule 
\endfirsthead
\caption{continued.} \\
\toprule\toprule
Type    & Magnitude range                & Color             & $N_{\rm LMC}$ & $Median_{\rm LMC}$ & $SD_{\rm LMC}$ & $N_{\rm SMC}$ & $Median_{\rm SMC}$ & $SD_{\rm SMC}$ \\
        & (mag)                          & (mag)             &         & (mag)        & (mag)        &         & (mag)        & (mag)        \\
\midrule
\endhead
\midrule
\endfoot
BSG   &  $M_{\rm G}\leq-6.5$               &  $BP-RP$      &  201   &  0.11       &  0.27       &  52    &  0.12       &  0.22 \\  
-     &  $-6.5<M_{\rm G}\leq-6.0$          &  -                &  181   &  0.012      &  0.25       &  57    &  0.12       &  0.23 \\
-     &  $-6.0<M_{\rm G}\leq-5.5$          &  -                &  343   &  -0.045     &  0.25       &  115   &  0.061      &  0.26 \\
-     &  $-5.5<M_{\rm G}\leq-5.0$          &  -                &  610   &  -0.046     &  0.29       &  178   &  0.032      &  0.29 \\
-     &  $-5.0<M_{\rm G}\leq-4.5$          &  -                &  809   &  0.011      &  0.29       &  270   &  0.082      &  0.25 \\
-     &  $-4.5<M_{\rm G}\leq-4.0$          &  -                &  825   &  0.16       &  0.26       &  262   &  0.24       &  0.22 \\
-     &  $-4.0<M_{\rm G}\leq-3.5$          &  -                &  730   &  0.36       &  0.21       &  152   &  0.42       &  0.16 \\
Total &  -                               &  -                &  3903  &  0.15       &  0.30       &  1104  &  0.16       &  0.27 \\
\midrule  
YSG   &  $M_{\rm G}\leq-6.5$               &  -                &  20    &  1.10       &  0.15       &  5     &  0.86       &  0.15 \\
-     &  $-6.5<M_{\rm G}\leq-6.0$          &  -                &  8     &  1.07       &  0.24       &  4     &  0.85       &  0.20 \\
-     &  $-6.0<M_{\rm G}\leq-5.5$          &  -                &  12    &  1.01       &  0.15       &  13    &  0.71       &  0.17 \\
-     &  $-5.5<M_{\rm G}\leq-5.0$          &  -                &  26    &  1.06       &  0.15       &  25    &  0.76       &  0.20 \\
-     &  $-5.0<M_{\rm G}\leq-4.5$          &  -                &  47    &  1.16       &  0.16       &  37    &  0.72       &  0.15 \\
-     &  $-4.5<M_{\rm G}\leq-4.0$          &  -                &  81    &  1.14       &  0.16       &  33    &  0.87       &  0.16 \\
-     &  $-4.0<M_{\rm G}\leq-3.5$          &  -                &  76    &  1.16       &  0.14       &  1     &  1.12       &  -    \\
Total &  -                               &  -                &  270   &  1.13       &  0.16       &  118   &  0.76       &  0.17 \\
\midrule  
RSG   &  $M_{\rm G}\leq-6.5$               &  -                &  112   &  2.21       &  0.35       &  72    &  1.89       &  0.22 \\
-     &  $-6.5<M_{\rm G}\leq-6.0$          &  -                &  209   &  2.18       &  0.34       &  110   &  1.78       &  0.14 \\
-     &  $-6.0<M_{\rm G}\leq-5.5$          &  -                &  311   &  1.99       &  0.25       &  112   &  1.69       &  0.18 \\
-     &  $-5.5<M_{\rm G}\leq-5.0$          &  -                &  348   &  1.88       &  0.28       &  165   &  1.62       &  0.36 \\
-     &  $-5.0<M_{\rm G}\leq-4.5$          &  -                &  460   &  1.78       &  0.24       &  234   &  1.60       &  0.29 \\
-     &  $-4.5<M_{\rm G}\leq-4.0$          &  -                &  577   &  1.70       &  0.21       &  305   &  1.53       &  0.29 \\
-     &  $-4.0<M_{\rm G}\leq-3.5$          &  -                &  453   &  1.65       &  0.23       &  109   &  1.56       &  0.28 \\
Total &  -                               &  -                &  2476  &  1.80       &  0.33       &  1107  &  1.63       &  0.29 \\
\midrule  
\midrule  
BSG   &  $M_{\rm V}\leq-6.5$               &  $B-V$        &  2     &  0.20       &  0.078      &  10    &  0.10       &  0.11 \\
-     &  $-6.5<M_{\rm V}\leq-6.0$          &  -                &  40    &  0.045      &  0.15       &  59    &  0.03       &  0.13 \\
-     &  $-6.0<M_{\rm V}\leq-5.5$          &  -                &  183   &  -0.020     &  0.16       &  105   &  -0.010     &  0.18 \\
-     &  $-5.5<M_{\rm V}\leq-5.0$          &  -                &  303   &  -0.030     &  0.16       &  170   &  -0.010     &  0.16 \\
-     &  $-5.0<M_{\rm V}\leq-4.5$          &  -                &  421   &  -0.01      &  0.19       &  197   &  0.02       &  0.17 \\
-     &  $-4.5<M_{\rm V}\leq-4.0$          &  -                &  418   &  0.060      &  0.17       &  197   &  0.12       &  0.16 \\
-     &  $-4.0<M_{\rm V}\leq-3.5$          &  -                &  274   &  0.20       &  0.19       &  95    &  0.24       &  0.13 \\
Total &  -                               &  -                &  1723  &  0.040      &  0.20       &  839   &  0.060      &  0.17 \\
\midrule  
YSG   &  $M_{\rm V}\leq6.5$                &  -                &  2     &  0.77       &  0.078      &  2     &  0.63       &  0.00 \\
-     &  $-6.5<M_{\rm V}\leq-6.0$          &  -                &  8     &  0.80       &  0.19       &  3     &  0.60       &  0.13 \\
-     &  $-6.0<M_{\rm V}\leq-5.5$          &  -                &  6     &  0.72       &  0.17       &  12    &  0.45       &  0.20 \\
-     &  $-5.5<M_{\rm V}\leq-5.0$          &  -                &  6     &  0.76       &  0.21       &  22    &  0.46       &  0.22 \\
-     &  $-5.0<M_{\rm V}\leq-4.5$          &  -                &  19    &  0.82       &  0.27       &  28    &  0.43       &  0.15 \\
-     &  $-4.5<M_{\rm V}\leq-4.0$          &  -                &  35    &  0.79       &  0.21       &  20    &  0.51       &  0.17 \\
-     &  $-4.0<M_{\rm V}\leq-3.5$          &  -                &  45    &  0.85       &  0.19       &  -     &  -          &  -    \\
Total &  -                               &  -                &  128   &  0.82       &  0.21       &  88    &  0.46       &  0.19 \\
\midrule  
RSG   &  $M_{\rm V}\leq6.5$                &  -                &  6     &  1.87       &  0.15       &  22    &  1.82       &  0.28 \\
-     &  $-6.5<M_{\rm V}\leq-6.0$          &  -                &  21    &  1.88       &  0.52       &  46    &  1.72       &  0.21 \\
-     &  $-6.0<M_{\rm V}\leq-5.5$          &  -                &  70    &  1.88       &  0.33       &  97    &  1.59       &  0.24 \\
-     &  $-5.5<M_{\rm V}\leq-5.0$          &  -                &  159   &  1.79       &  0.33       &  112   &  1.50       &  0.28 \\
-     &  $-5.0<M_{\rm V}\leq-4.5$          &  -                &  187   &  1.70       &  0.29       &  143   &  1.38       &  0.27 \\
-     &  $-4.5<M_{\rm V}\leq-4.0$          &  -                &  261   &  1.60       &  0.28       &  193   &  1.32       &  0.30 \\
-     &  $-4.0<M_{\rm V}\leq-3.5$          &  -                &  301   &  1.47       &  0.23       &  205   &  1.25       &  0.26 \\
Total &  -                               &  -                &  1194  &  1.57       &  0.29       &  835   &  1.39       &  0.30 \\
\midrule  
\midrule  
BSG   &  $M_{K_{\rm S}}\leq-7.0$           &  $J-K_{\rm S}$      &  162   &  0.15       &  0.31       &  45    &  0.13       &  0.27 \\
-     &  $-7.0<M_{K_{\rm S}}\leq-6.5$      &  -                &  138   &  0.093      &  0.24       &  48    &  0.14       &  0.26 \\
-     &  $-6.5<M_{K_{\rm S}}\leq-6.0$      &  -                &  255   &  0.089      &  0.16       &  77    &  0.17       &  0.13 \\
-     &  $-6.0<M_{K_{\rm S}}\leq-5.5$      &  -                &  380   &  0.073      &  0.18       &  157   &  0.13       &  0.18 \\
-     &  $-5.5<M_{K_{\rm S}}\leq-5.0$      &  -                &  740   &  0.083      &  0.17       &  263   &  0.16       &  0.17 \\
-     &  $-5.0<M_{K_{\rm S}}\leq-4.5$      &  -                &  1110  &  0.093      &  0.16       &  344   &  0.12       &  0.16 \\
-     &  $-4.5<M_{K_{\rm S}}\leq-4.0$      &  -                &  990   &  0.069      &  0.16       &  167   &  0.075      &  0.16 \\
Total &  -                               &  -                &  3898  &  0.083      &  0.18       &  1104  &  0.12       &  0.18 \\
\midrule  
YSG   &  $M_{K_{\rm S}}\leq-7.0$           &  -                &  74    &  0.56       &  0.22       &  23    &  0.43       &  0.12 \\
-     &  $-7.0<M_{K_{\rm S}}\leq-6.5$      &  -                &  47    &  0.57       &  0.15       &  20    &  0.37       &  0.13 \\
-     &  $-6.5<M_{K_{\rm S}}\leq-6.0$      &  -                &  73    &  0.61       &  0.13       &  43    &  0.36       &  0.089 \\
-     &  $-6.0<M_{K_{\rm S}}\leq-5.5$      &  -                &  64    &  0.49       &  0.086      &  28    &  0.34       &  0.068 \\
Total &  -                               &  -                &  270   &  0.54       &  0.16       &  118   &  0.36       &  0.10 \\
\midrule  
RSG   &  $M_{K_{\rm S}}\leq-9.5$           &  -                &  379   &  1.12       &  0.13       &  95    &  0.99       &  0.15 \\
-     &  $-9.5<M_{K_{\rm S}}\leq-9.0$      &  -                &  219   &  1.07       &  0.075      &  101   &  0.96       &  0.11 \\
-     &  $-9.0<M_{K_{\rm S}}\leq-8.5$      &  -                &  273   &  1.04       &  0.082      &  114   &  0.93       &  0.12 \\
-     &  $-8.5<M_{K_{\rm S}}\leq-8.0$      &  -                &  297   &  1.01       &  0.072      &  137   &  0.91       &  0.12 \\
-     &  $-8.0<M_{K_{\rm S}}\leq-7.5$      &  -                &  352   &  0.98       &  0.077      &  190   &  0.88       &  0.10 \\
-     &  $-7.5<M_{K_{\rm S}}\leq-7.0$      &  -                &  484   &  0.94       &  0.078      &  204   &  0.86       &  0.11 \\
-     &  $-7.0<M_{K_{\rm S}}\leq-6.5$      &  -                &  412   &  0.89       &  0.070      &  230   &  0.81       &  0.075 \\
Total &  -                               &  -                &  2476  &  0.99       &  0.12       &  1107  &  0.88       &  0.13 \\
\midrule  
\midrule  
BSG   &  $M_{\rm IRAC2}\leq-7.0$         &  $IRAC1-IRAC2$    &  196   &  0.073      &  0.20       &  54    &  0.070      &  0.20 \\
-     &  $-7.0<M_{\rm IRAC2}\leq-6.5$    &  -                &  144   &  0.046      &  0.060      &  53    &  0.043      &  0.055 \\
-     &  $-6.5<M_{\rm IRAC2}\leq-6.0$    &  -                &  279   &  0.046      &  0.098      &  95    &  0.041      &  0.085 \\
-     &  $-6.0<M_{\rm IRAC2}\leq-5.5$    &  -                &  442   &  0.046      &  0.099      &  180   &  0.041      &  0.092 \\
-     &  $-5.5<M_{\rm IRAC2}\leq-5.0$    &  -                &  817   &  0.039      &  0.082      &  280   &  0.040      &  0.092 \\
-     &  $-5.0<M_{\rm IRAC2}\leq-4.5$    &  -                &  1046  &  0.031      &  0.071      &  317   &  0.033      &  0.066 \\
-     &  $-4.5<M_{\rm IRAC2}\leq-4.0$    &  -                &  793   &  0.027      &  0.064      &  113   &  0.020      &  0.037 \\
Total &  -                               &  -                &  3739  &  0.036      &  0.092      &  1069  &  0.038      &  0.092 \\
\midrule  
YSG   &  $M_{\rm IRAC2}\leq-7.0$         &  -                &  80    &  0.0089     &  0.16       &  26    &  0.033      &  0.061 \\
-     &  $-7.0<M_{\rm IRAC2}\leq-6.5$    &  -                &  45    &  -0.018     &  0.044      &  22    &  0.036      &  0.030 \\
-     &  $-6.5<M_{\rm IRAC2}\leq-6.0$    &  -                &  81    &  -0.031     &  0.066      &  49    &  0.037      &  0.034 \\
-     &  $-6.0<M_{\rm IRAC2}\leq-5.5$    &  -                &  56    &  0.0060     &  0.069      &  20    &  0.038      &  0.023 \\
Total &  -                               &  -                &  258   &  -0.013     &  0.10       &  116   &  0.035      &  0.039 \\
\midrule  
RSG   &  $M_{\rm IRAC2}\leq-9.5$         &  -                &  265   &  -0.097     &  0.19       &  99    &  -0.010     &  0.18 \\
-     &  $-9.5<M_{\rm IRAC2}\leq-9.0$    &  -                &  228   &  -0.090     &  0.21       &  95    &  -0.087     &  0.070 \\
-     &  $-9.0<M_{\rm IRAC2}\leq-8.5$    &  -                &  275   &  -0.15      &  0.12       &  109   &  -0.095     &  0.043 \\
-     &  $-8.5<M_{\rm IRAC2}\leq-8.0$    &  -                &  295   &  -0.15      &  0.048      &  136   &  -0.093     &  0.045 \\
-     &  $-8.0<M_{\rm IRAC2}\leq-7.5$    &  -                &  354   &  -0.14      &  0.055      &  202   &  -0.083     &  0.040 \\
-     &  $-7.5<M_{\rm IRAC2}\leq-7.0$    &  -                &  475   &  -0.12      &  0.038      &  194   &  -0.068     &  0.038 \\
-     &  $-7.0<M_{\rm IRAC2}\leq-6.5$    &  -                &  424   &  -0.11      &  0.039      &  223   &  -0.062     &  0.051 \\
Total &  -                               &  -                &  2267  &  -0.13      &  0.10       &  1080  &  -0.075     &  0.074 \\
\midrule  
\midrule  
BSG   &  $M_{\rm IRAC4}\leq-7.0$         &  $J-IRAC4$        &  215   &  0.47       &  1.19       &  66    &  0.39       &  1.58 \\
-     &  $-7.0<M_{\rm IRAC4}\leq-6.5$    &  -                &  195   &  0.33       &  0.62       &  67    &  0.31       &  0.62 \\
-     &  $-6.5<M_{\rm IRAC4}\leq-6.0$    &  -                &  321   &  0.34       &  0.59       &  124   &  0.34       &  0.65 \\
-     &  $-6.0<M_{\rm IRAC4}\leq-5.5$    &  -                &  486   &  0.29       &  0.55       &  195   &  0.30       &  0.51 \\
-     &  $-5.5<M_{\rm IRAC4}\leq-5.0$    &  -                &  727   &  0.22       &  0.42       &  221   &  0.21       &  0.34 \\
-     &  $-5.0<M_{\rm IRAC4}\leq-4.5$    &  -                &  742   &  0.14       &  0.27       &  207   &  0.11       &  0.20 \\
-     &  $-4.5<M_{\rm IRAC4}\leq-4.0$    &  -                &  545   &  -0.00053   &  0.26       &  111   &  -0.055     &  0.24 \\
Total &  -                               &  -                &  3343  &  0.20       &  0.58       &  997   &  0.21       &  0.65 \\
\midrule 
YSG   &  $M_{\rm IRAC4}\leq-7.0$         &  -                &  78    &  0.77       &  0.42       &  28    &  0.62       &  0.23 \\
-     &  $-7.0<M_{\rm IRAC4}\leq-6.5$    &  -                &  45    &  0.73       &  0.19       &  26    &  0.49       &  0.15 \\
-     &  $-6.5<M_{\rm IRAC4}\leq-6.0$    &  -                &  81    &  0.71       &  0.17       &  43    &  0.50       &  0.15 \\
-     &  $-6.0<M_{\rm IRAC4}\leq-5.5$    &  -                &  46    &  0.61       &  0.15       &  19    &  0.47       &  0.11 \\
Total &  -                               &  -                &  255   &  0.71       &  0.29       &  117   &  0.50       &  0.18 \\
\midrule 
RSG   &  $M_{\rm IRAC4}\leq-9.5$         &  -                &  435   &  1.52       &  0.46       &  143   &  1.26       &  0.49 \\
-     &  $-9.5<M_{\rm IRAC4}\leq-9.0$    &  -                &  211   &  1.22       &  0.12       &  98    &  1.13       &  0.20 \\
-     &  $-9.0<M_{\rm IRAC4}\leq-8.5$    &  -                &  238   &  1.19       &  0.10       &  113   &  1.10       &  0.18 \\
-     &  $-8.5<M_{\rm IRAC4}\leq-8.0$    &  -                &  308   &  1.14       &  0.10       &  149   &  1.06       &  0.13 \\
-     &  $-8.0<M_{\rm IRAC4}\leq-7.5$    &  -                &  366   &  1.10       &  0.086      &  193   &  1.02       &  0.13 \\
-     &  $-7.5<M_{\rm IRAC4}\leq-7.0$    &  -                &  482   &  1.06       &  0.081      &  202   &  0.98       &  0.12 \\
-     &  $-7.0<M_{\rm IRAC4}\leq-6.5$    &  -                &  308   &  0.99       &  0.075      &  186   &  0.93       &  0.076 \\
Total &  -                               &  -                &  2375  &  1.13       &  0.313      &  1094  &  1.04       &  0.27 \\
\midrule 
\midrule 
BSG   &  $M_{\rm MIPS24}\leq-12.0$       &  $K_{\rm S}-MIPS24$&  25    &  7.68       &  1.60       &  6     &  6.46       &  1.72 \\
-     &  $-10.0<M_{\rm MIPS24}\leq-12.0$ &  -                &  35    &  5.03       &  2.29       &  13    &  5.31       &  1.34 \\
-     &  $-8.0<M_{\rm MIPS24}\leq-10.0$  &  -                &  67    &  3.04       &  1.67       &  26    &  3.25       &  1.42 \\
Total &  -                               &  -                &  133   &  4.21       &  2.57       &  47    &  3.74       &  2.01 \\
\midrule 
RSG   &  $M_{\rm MIPS24}\leq-11.0$       &  -                &  259   &  2.46       &  0.87       &  37    &  2.13       &  1.49 \\
-     &  $-11.0<M_{\rm MIPS24}\leq-10.5$ &  -                &  62    &  1.09       &  0.25       &  13    &  0.63       &  0.29 \\
-     &  $-10.5<M_{\rm MIPS24}\leq-10.0$ &  -                &  80    &  0.77       &  0.24       &  36    &  0.50       &  0.33 \\
-     &  $-10.0<M_{\rm MIPS24}\leq-9.5$  &  -                &  82    &  0.52       &  0.21       &  66    &  0.39       &  0.27 \\
-     &  $-9.5<M_{\rm MIPS24}\leq-9.0$   &  -                &  151   &  0.34       &  0.17       &  71    &  0.31       &  0.20 \\
-     &  $-9.0<M_{\rm MIPS24}\leq-8.5$   &  -                &  160   &  0.28       &  0.23       &  70    &  0.28       &  0.29 \\
-     &  $-8.5<M_{\rm MIPS24}\leq-8.0$   &  -                &  148   &  0.28       &  0.19       &  40    &  0.36       &  0.22 \\
Total &  -                               &  -                &  965   &  0.52       &  1.05       &  337   &  0.39       &  0.81 \\
\midrule  
\end{longtable}
\tablefoot{
Since there may be a few targets out of the magnitude ranges, the total number may be different from the sum of the numbers of all magnitude ranges.
}
}

\clearpage

\end{CJK*}

\end{document}